\documentclass[fleqn,usenatbib]{mnras}

\usepackage{newtxtext,newtxmath}

\usepackage[T1]{fontenc}
\usepackage{ae,aecompl}

\usepackage{graphicx}	
\usepackage{amsmath}	
\usepackage{amssymb}	
\usepackage{gensymb}    
\usepackage{rotating}   
\usepackage{multirow}

\newcommand {\beq} {\begin{equation}}
\newcommand {\eeq} {\end{equation}}
\newcommand {\bdm} {\begin{displaymath}}
\newcommand {\edm} {\end{displaymath}}

\title[Internal shocks in microquasar jets]{Internal shocks in microquasar jets\\ with a continuous Lorentz factor modulation}

\author[P.~Pjanka and~J.~M.~Stone]{
Patryk Pjanka,$^{1}$\thanks{E-mail: \href{ppjanka@princeton.edu}{ppjanka@princeton.edu}}
and James M. Stone$^{1}$
\\
$^{1}$Department of Astrophysical Sciences, Princeton University, 4 Ivy Lane, Princeton, NJ 08544, USA
}

\date{Accepted XXX. Received YYY; in original form ZZZ}

\pubyear{2018}

\begin{document}
\label{firstpage}
\pagerange{\pageref{firstpage}--\pageref{lastpage}}
\maketitle

\begin{abstract}
We perform relativistic hydrodynamic simulations of internal shocks formed in microquasar jets by continuous variation of the bulk Lorentz factor, in order to investigate the internal shock model. We consider one-, two-, and flicker noise 20-mode variability.
We observe emergence of a forward-reverse shock structure for each peak of the Lorentz factor modulation. The high pressure in the shocked layer launches powerful outflows perpendicular to the jet beam into the ambient medium. These outflows dominate the details of the jet's kinetic energy thermalization. They are responsible for mixing between the jet and surrounding medium and generate powerful shocks in the latter. 
These results do not concur with the popular picture of well-defined internal shells depositing energy as they collide within the confines of the jet, in fact collisions between internal shells themselves are quite rare in our continuous formulation of the problem. 
For each of our simulations, we calculate the internal energy deposited in the system, the ``efficiency'' of this deposition (defined as the ratio of internal to total flow energy), and the maximum temperature reached in order to make connections to emission mechanisms. We probe the dependence of these diagnostics on the Lorentz factor variation amplitudes, modulation frequencies, as well as the initial density ratio between the jet and the ambient medium.
\end{abstract}




\section{Introduction}


Relativistic jets are one of the most energetic phenomena observed in nature. While, for simplicity, they are often treated as stationary flows (especially in the low/hard states of X-ray binaries; e.g., \citealt{1979BlandfordKonigl, 2006Bosch-Ramon, 2014Zdziarski}), it seems unlikely that they would be so calm in reality. Instabilities in the accretion disk can modulate the mass inflow at the ejection site \citep[see][]{2015Tchekhovskoy}, while instabilities within the jet itself can modulate the outflow \citep[e.g.,][]{2011Granot}. Inhomogeneous jet models have therefore been proposed to explain a variety of phenomena observed in jetted systems. In most models to date the jet is considered to be composed of a number of discrete shells with varying density, speed, and/or magnetic field strength. These shells then collide forming internal shocks (hence, ``internal shock models''), which redistribute energy between the bulk motion, magnetic field, thermal and non-thermal particles within the jet.


Initially, most of the interest in the internal shock model was sparked by its ability to relax constraints on conditions in gamma-ray bursts (GRBs; e.g., \citealt{1994Rees, 1994Paczynski, 1998Daigne, 2011Granot, 2012GranotA, 2012GranotB}). But internal shock models are also quite promising when it comes to understanding relativistic jets in X-ray binaries (XRBs) and active galactic nuclei (AGN). In their hard states, these jets are often well described by the conical jet model of \cite{1979BlandfordKonigl}. One of its main features, a flat radio spectrum, is attributed to the electron energy distribution remaining constant along the jet, which requires a reheating mechanism for non-thermal particles. In addition, emission models of XRB/AGN jets indicate that the magnetization $\sigma_B$ in radio-emitting parts of the jet is $<<1$, while analytical considerations and simulations of magnetically-driven jets ejection require $\sigma_B \sim 1$ at the ejection site \citep{2010Lyubarsky, 2009Tchekhovskoy, 2015Zdziarski}. The two processes currently investigated to resolve the two issues are magnetic reconnection and the presence of internal shocks \citep[see][and references therein]{2011Komissarov, 2015Zdziarski}. Here, we will focus on the second mechanism.



Much of the work concerning the internal shock model in microquasars has been dedicated to understanding a single collision between two shells in an otherwise homogeneous relativistic jet. \cite{2004Mimica} performed 2D relativistic hydrodynamic simulations of a collision between two dense shells moving in a rarified pressurized external medium. They provided a careful treatment of the pre-collision evolution of the shells (including analytical estimates of the resulting profiles of hydrodynamical parameters) and a thorough analysis of how the collision dynamics is affected by varying parameters of the shells. With a clever method of injecting non-thermal emitting particles at the shocks as passive fields in their MHD code, they calculated X-ray spectra and light curves of their models. They find emission from the collisions to be extremely variable, as the electron cooling time they find is short compared to the shell collision timescales. The total radiated energy and light curves of \cite{2004Mimica} are consistent with flaring BL~Lac sources and the radiative efficiency of their model is $1\%$. Since they find lateral expansion in internal shocks to be unimportant, \cite{2005Mimica} perform a study of internal shocks in BL~Lac flares using 1D hydrodynamical simulations. They find a correlation between hard-soft peak emission lags and initial rest mass density of the colliding shells. A similar case was studied analytically by \cite{2010Bottcher}. They designed a one-zone emission model for two colliding unmagnetized shells with strict particle acceleration treatment, as well as light-travel and shock movement delays taken into account in the light curve calculation. They provide a detailed analysis of the influence of various parameters of the colliding shells on observed spectra and variability of blazars. Their results successfully reproduced the transition from BL~Lacs to FSRQs \citep{1998Fossati, 1998Ghisellini}.


Quasi-analytical models of multiple-shell systems have also been investigated. \cite{2001Spada} considered hierarchical merging of unmagnetized shells ejected from the central engine with a flat distribution of parameters selected from ranges of observationally reasonable values. Analytical hydrodynamic prescriptions for the results of each collision (emission, particle heating) were used. They successfully reproduced general features of the blazar 3C~279. \cite{2010Jamil} expanded the model of \cite{2001Spada} and applied it to single shell propagation, two-shell collision, and multiple shell models. They point out that a non-zero internal energy of the shells is necessary for the shells to reproduce flat synchrotron spectrum in the presence of adiabatic losses. Their models successfully reproduce the scaling of the high-frequency spectral break with jet power, $\nu_b \propto P_j^{\sim 0.6\textrm{---}0.7}$, in accord with observations and analytical models \citep{1995Falcke, 2003Heinz, 2003Markoff}.

\cite{2013Malzac, 2014Malzac} conducted a detailed treatment of variability and emission in an internal shock model with multiple colliding shells using analytical hydrodynamic prescriptions for the results of each collision. Analytical predictions of the observed spectrum based on Fourier analysis of the shell properties were compared to numerical simulations of a finite number of distinct shells. Cases of Dirac delta, flicker noise and power-law power spectral densities (PSDs) of the shell bulk Lorentz factors were analyzed in detail. \cite{2013Malzac} concluded that XRB jet properties can be reproduced by an internal shock model with a flicker-noise PSD of the bulk Lorentz factors of injected shells. He notes that this type of variability is consistent with PSD of X-ray emission from XRBs \citep{2002Reig, 2003Reig, 2005Gilfanov, 2010Gilfanov}, and hence, the accretion rate variability. \cite{2014Malzac} also observed that the internal shock models predict considerable time-dependent variability in XRB jet emission, in agreement with observations.

Modulation of the jet flow parameters does not have to manifest as multiple discrete shells. Some AGN outbursts have been observed to be accompanied by a single shock/shell propagation through the system's jet (e.g., \citealt{1978Rees, 2015Meyer}). Following this clue, \cite{2000Kaiser} considered a single fast shell propagating through a homogeneous jet. They found that particle acceleration by the moving shock and subsequent cooling can explain time-variable spectra of outbursts in XRBs.


In this work, we aim to expand the considerations of multiple colliding unmagnetized shells into the continuous case. Instead of treating the ejected shells as discrete cone sections of plasma moving with different Lorentz factors, we simulate a plasma flow with continuously varying Lorentz factors. This allows us to observe the emergence of shell-like structures in the jet (through shock steepening) and their subsequent evolution.

In Section~\ref{sect:methods} we shortly introduce the astrophysical magnetohydrodynamics (MHD) code \texttt{Athena++} (Sect.~\ref{sect:athenapp}), the implementation of our problem (Sect.~\ref{sect:numericalSetup}), and describe how fluctuations with a custom Lorentz factor PSD were introduced (Sect.~\ref{sect:shellInjection}). We describe the results of our simulations for a single shell, two-shell and multiple-shell models in Section~\ref{sect:results}. Finally, in Section~\ref{sect:discussion}, we discuss and summarize our findings.

\section{Methods}\label{sect:methods}

\subsection{\texttt{Athena++}}\label{sect:athenapp}

We performed our simulations using the astrophysical magnetohydrodynamics (MHD) code \texttt{Athena++}\footnote{\texttt{Athena++} is publicly available through the project website \url{http://princetonuniversity.github.io/athena/}}. It is a finite-volume 3D MHD Godunov code, utilizing staggered-mesh constrained transport method to enforce the divergence-free constraint on the magnetic field, and supports both special and general relativity in stationary spacetimes \citep{Athena++-WhiteStone2016}. The numerical algorithms in \texttt{Athena++} are described in previous method papers: \cite{Athena-StoneGardiner2008} and \cite{Athena-BeckwithStone2011}.

\subsection{Numerical setup}\label{sect:numericalSetup}

\subsubsection{Boundary conditions, resolution, and grid description}\label{sect:numericalSetup:boundary}

We simulate a section of the jet environment up to $5$ jet radii from the jet axis, far enough downstream that the system can be approximated as a hydrodynamical parallel flow. The simulations are, in essence, two-dimensional and cover a radial slice of the system, with the $r = 0$ axis corresponding to the jet axis. In order to allow for adiabatic cooling of the matter moving away from the jet axis, we set up the simulations as three-dimensional in a cylindrical coordinate system $(r, \phi, z)$, with the azimuthal dimension collapsed to only 4 zones covering $1$~radian in the $\phi$ angle. In the radial dimension $r$, the simulated domain covers a radius of $5$~jet radii $r_{\rm jet}$ at a resolution of $256$ zones. In the direction of the jet axis $z$, the size of the box is set to cover at least $5$~wavelengths of the lowest-frequency modulation component injected (see Sect.~\ref{sect:numericalSetup:shellinjection}). The resolution in this direction is either $1024$~zones or such that the shortest-wavelength sinusoidal component of the modulation is covered with at least $18$~zones, whichever is larger.

The boundary conditions in the direction of the jet flow (minimal and maximal $z$, boundaries perpendicular to the jet axis) are that of uninhibited outflows/inflows. The boundaries at constant $\phi$ are set as reflective, so that we avoid buildup of azimuthal waves while allowing correct adiabatic behavior of gas flowing away from the jet axis. We stress that we do not attempt to conduct 3D simulations of the jet and only introduce the $\phi$ dimension to correctly reproduce the behavior of a cylindrical system in our 2D approach. The ``jet side'' boundary at $5r_{\rm jet}$ is set to allow free outflows/inflows, as this boundary does not convey any physical meaning and only sets an arbitrary limit to the region we are interested in, the immediate vicinity of the original jet. The inner radial boundary at $r=0$ is treated as reflective.

\subsubsection{Initialization}\label{sect:numericalSetup:initialization}

We start our simulations with a stationary cylindrical laminar relativistic flow within a static ambient medium. In order to resolve the flow, the physical parameters of the medium change with radius smoothly following the ``jet fraction'' in the medium:
\beq f(r) = \frac{1}{2} - \frac{\arctan\left((r - r_{\rm jet})/\Delta r_{\rm smooth}\right)}{\pi}, \eeq
where the smoothing parameter $\Delta r_{\rm smooth}$ was set to $0.01r_{\rm jet}$. The hydrodynamic properties of the medium were initialized as follows:
\begin{align} \begin{split}
\rho(r) &= f(r)\rho_{\rm jet} + (1-f(r))\rho_{\rm amb}, \\
\Gamma(r) &= f(r)\Gamma_{\rm jet} + (1-f(r))\Gamma_{\rm amb}, \\
v_z(r) &= \sqrt{1-\Gamma^{-2}(r)}, \\
v_r(r) &= v_{\phi}(r) = 0 \\
P(r) &= P_{\rm jet} = P_{\rm amb} = 5.56\times 10^{-2}\textrm{ sim.u.}, \\ 
H(r) &= 1+\frac{\gamma P}{(\gamma-1)\rho},
\end{split}\end{align}
where $\rho$ denotes density in the fluid's rest frame; $\Gamma$ -- the fluid's Lorentz factor as seen in the LAB frame; $v_r, v_{\phi}, v_z$ stand for velocity components for the LAB frame in $r, \phi$, and $z$ directions, respectively, in units of the speed of light; $P$ is the fluid pressure in its rest frame; $H$ is the rest-frame enthalpy of the fluid; and $\gamma$ is the adiabatic index. We assume that the pressure in our system is dominated by the pressure of relativistic electrons and, hence, set $\gamma = 4/3$. Subscripts ``jet'' and ``amb'' are added to values corresponding to the jet and the ambient medium, respectively. These values are the parameters of our model.

\subsubsection{Shell pattern injection}\label{sect:numericalSetup:shellinjection}

The goal of our work is to study hydro- and thermodynamical evolution of internal shocks in relativistic jets. To achieve this, we modulate the Lorentz factor of matter flowing into the simulated domain over time. To aid parametrization of this modulation, we represent it as a sum of sinusoidal components of the variation of the jet Lorentz factor $\Gamma_{\rm jet}$ around a default value, $\Gamma_{\rm jet, 0}$. For each injected pattern, $\Gamma_{\rm jet, 0}$ is set to the value ensuring that the minimal Lorentz factor reached is $1$. We refer to each single sinusoidal component as a ``modulation component'' and parametrize it by $\Gamma_{\rm max}$, the (absolute) maximal Lorentz factor reached in the pattern if it was the only component injected, and $\omega$, the frequency of modulation. As a result, the Lorentz factor of the fluid at the jet boundary where matter flows into the simulation box ($z=0$) can be expressed as:
\beq \begin{split} \Gamma(r, z=0, t) = f(r)\left( \Gamma_{\rm jet,0} + \sum_{i=1}^N{(\Gamma_{{\rm jet},i}-\Gamma_{\rm jet,0})\sin(\omega_it)} \right)\\ + (1-f(r))\Gamma_{\rm amb}. \end{split}\eeq

\subsubsection{Shell pattern injection with a flicker-noise Power Spectral Density}\label{sect:shellInjection}

In addition to investigating the propagation of single- and two-component patterns, we also test a more complex Power Spectral Density (PSD) of the jet Lorentz factor variation. As a case study, we have chosen the flicker-noise PSD, which was found by \cite{2014Malzac} to reliably reproduce the flat radio spectrum of jets at low frequencies and corresponds to the character of variability in some AGN. In our case, the PSD was approximated by $20$ sinusoidal components ($\Gamma_{{\rm max},i}$, $\omega_i$):
\beq \textrm{PSD}(\omega) = \sum_{i=0}^{19}\Gamma_{{\rm max},i}^2\delta(\omega-\omega_i), \eeq
where $\delta(\omega)$ is the Dirac delta. In order for the PSD to describe a flicker noise, i.e., $\textrm{PSD} \propto \omega^{-1}$, we set $\Gamma_{{\rm max},i} \propto \omega_i^{-1/2}$. To find a representation, we first draw $\omega_i$ from a flat distribution in the $[\omega_{\rm min}, \omega_{\rm max}] = [0.1,10]$ range (in sim.u., limited by our numerical resources). Then the amplitude of the pattern is set to be
\beq \Delta\Gamma_i = 5\times\left(\frac{\omega_i}{\omega_{\rm min}}\right)^{-1/2} \in [0.5, 5.0]. \eeq
The oscillation's phase is chosen by delaying the injection of each component by $\Delta t$ drawn from a flat distribution in the range $[0,2\pi/\omega_i)$. We repeat the process for each of the $20$ components independently. Then, the composite pattern is assembled and its minimum within 5 wavelengths of the lowest-frequency component is calculated. This minimum, multiplied by $-1$ and increased by $1.01$, becomes the average jet Lorentz factor, $\Gamma_{\rm max, 0}$. Its value is added to each $\Delta\Gamma_i$, yielding $\Gamma_{\rm max,i} = \Delta\Gamma_i + \Gamma_{\rm max, 0}$.

To calculate our diagnostics (see Sect.~\ref{sect:diagnostics}) and correctly initialize a simulation (Sect.~\ref{sect:numericalSetup:initialization}), we require wavelengths of our pattern components. We approximate them as
\beq \lambda_{\rm i} = \sqrt{1-\Gamma_{\rm max, i}^{-2}}\frac{2\pi c}{\omega_i}. \eeq

\subsubsection{Testing for numerical dissipation at shock positions}

In order to ensure that we adequately capture shocks in our numerical approach, we examined the widths of shocks generated in a simplified version of our model. We performed one-dimensional simulations of the jet axis, with all the simulation setup following that of our main set of simulations except for the outer radial boundary condition, which was set as reflective. A single-component sinusoidal pattern of Lorentz factor modulation was used. The simulation was run until $t=50$~sim.u., at which point the shocks that formed were located and measured. To measure the widths, we fit a Gaussian to the (finite-difference) derivative of the fluid Lorentz factor over $z$ at each shock position. The variances of these fits were then averaged over all shocks in the final snapshot of each simulation. We performed this procedure for three different amplitudes of the Lorentz factor modulation, $\Gamma_{\rm max}\in {2,5,25}$, and at resolutions of 256, 512, 1024, 2048, and 4096 cells per $50$~sim.u (note that the default resolution is $1024/50$~sim.u., Sect.~\ref{sect:numericalSetup:boundary}). In all our simulations the width of the shocks remained in the range of $2.0$-$3.5$ cell widths, with no discernible resolution dependence, and the shocks for $\Gamma_{\rm max}=25$ being slightly wider. This stability in shock capturing proves that our numerical scheme is appropriate for investigation of shocks at high Lorentz factors.

\subsubsection{Scaling simulation units to Cyg~X-1}\label{sect:numericalSetup:sim2phys}

In order to guide the analysis and aid interpretation of our results, we have adopted a specific case to give physical dimensions to our simulation units. We choose Model~1 of the Cyg~X-1 system built by \cite{2014Zdziarski} to relate our simulation to physical quantities. We stress that we in no way intend to rigorously model Cyg~X-1. Our analysis involves a wide range of Lorentz factor values and variability frequencies, many of which will not correspond to conditions found in this system. However, we find it helpful in interpretation of our results to have them anchored in physical units, even if these do not faithfully reproduce the same source in different cases.

The jet opening angle in the hard state of Cyg~X-1 is known to be $\Theta_j \lesssim 2\degree$ \citep{Stirling2001}. We place the simulated box at $800$ gravitational radii ($r_g = 2.36\times 10^6$~cm, \citealt{Ziolkowski2014}) from the black hole, which gives a jet radius of $r_{\rm jet} = \Theta_j\times800r_g \simeq 6.59\times10^7$~cm. This sets both the length and time dimensions of our results:
\begin{align} 
&1\textrm{ length sim. u.} = r_{\rm jet} \simeq 6.59\times10^7\textrm{ cm}, \\
&1\textrm{ time sim. u.} = r_{\rm jet}/c \simeq 2.20\textrm{ ms}. \end{align}
To normalize the density, we use the jet power estimation from Model~1 of \cite{2014Zdziarski}:
\beq P_i \sim \Gamma_j^2\rho_{\rm jet} c^2v_{\rm jet}c\times \pi r_{\rm jet}^2 \simeq 10^{36.6}\textrm{ erg}/\textrm{s}. \eeq
Since in our simulations $\Gamma_j$ is highly variable, we use the average Lorentz factor $\Gamma_{\rm jet, 0}$ and the corresponding velocity $v_{\rm jet, 0} = \sqrt{1-1/\Gamma_{\rm jet, 0}^2}$. We obtain normalizations for density and pressure:
\beq \begin{split} 1\textrm{ density sim. u.} = \rho_{\rm amb} = \left(\frac{\rho_{\rm jet}}{\rho_{\rm amb}}\right)^{-1}\frac{P_i}{\Gamma_{\rm jet, 0}^2 v_{\rm jet,0}c^3\times \pi r_{\rm jet}^2}\\ \simeq 1.08\times 10^{-11}\frac{\textrm{g}}{\textrm{cm}^3}\times \left(\frac{\rho_{\rm jet}}{\rho_{\rm amb}}\right)^{-1}\frac{1}{\Gamma_{\rm jet, 0}^2v_{\rm jet,0}}, \end{split} \eeq
\beq \begin{split} 1\textrm{ pressure sim. u.}& = \rho_{\rm amb}c^2 \\
\simeq & 9.73\times 10^{8}\frac{\textrm{erg}}{\textrm{cm}^3}\times \left(\frac{\rho_{\rm jet}}{\rho_{\rm amb}}\right)^{-1}\frac{1}{\Gamma_{\rm jet, 0}^2v_{\rm jet,0}}, \end{split}\eeq
both of which depend on the average Lorentz factor of the given simulation.

\subsection{Diagnostics}\label{sect:diagnostics}

A lot can be learned about evolution of the internal shocks in our models from examination of the distribution of density, pressure, and fluid velocity in our simulations. We augment this set by introducing additional local and global diagnostics.

The first local diagnostic we consider is the temperature of the fluid. We derive it by assuming the fluid to be a fully ionized ideal hydrogen gas, with mass supplied by non-relativistic protons and pressure provided by relativistic gas of electrons. In such a case, the pressure is given by:
\beq P = \frac{\rho}{\mu m_p}kT, \eeq
where $\mu=1$ is the number of pressure-exerting particles per proton mass ($m_p$) of the fluid. This relation straightforwardly results in a measure of temperature $T$. We stress that the jet fluid is not expected to be in thermodynamic equilibrium -- the energy deposited in the fluid will lead to particle acceleration and emergence of a non-thermal distribution. The temperatures we report should only be treated as a measure of internal energy per particle, indicating plausibility of particle acceleration and enhanced radiation in a given region of the system.

To have a basis for a global comparison of different models, we also implement four global diagnostics.

The first and simplest is the maximum temperature reached within the simulation domain. It carries information on the ability of the internal-shock-related structures to accelerate particles, as it captures regions of high density of internal energy, usually corresponding to powerful shocks.

The second characteristic is the internal energy deposited in the system. Since we investigate Lorentz factor modulation of different sizes, we report the internal energy deposition per unit length along the jet axis. We divide the simulation domain into cylindrical sections, each with a length equal to the wavelength of the lowest-frequency modulation component injected, $\lambda_{\rm max}$. In each cylindrical section, for each moment in time, we integrate the internal energy over the volume within $3$ jet radii from the jet axis. While doing so, we account for the fact that the simulation domain is only a slice of the full cylindrical system (i.e., correct by a factor of $2\pi/\Delta\phi$). The internal energy within each section becomes close to periodical after the second wavelength of the pattern passes through it (see Fig.~\ref{fig:qualitative} and Sect.~\ref{sect:qualitative}), so we consider the internal energy averaged over time after this moment to be the average deposited energy within the given section. Interestingly, as we can see, e.g., in Fig.~\ref{fig:res_nshell}, this average deposited energy does not always saturate at a specific point downstream from the injection site. Instead, there appears to be a moment of maximal internal energy content, after which the fluid decompresses and returns some of the internal energy into the bulk kinetic energy. In a real jet, the internal energy may be used at the point of maximum to transfer heat into other forms of energy (magnetic field, particle acceleration, etc.), so it seems unlikely that the fluid would follow such a simple decompression. We therefore select the average internal energy content for the cylindrical section where it is maximal, and report its value divided by the length of the section as our diagnostic.

The disadvantage of using the internal energy deposition per unit length is that it is very sensitive to the Lorentz factors of the injected pattern. Since, for a highly relativistic motion, the internal and kinetic energy depend on the Lorentz factor of the fluid in a similar way, and the internal energy is generated through changes in the kinetic energy; we find it useful to define a ``deposition efficiency'' $\eta_{\rm dep}$ as the ratio of the internal to total (i.e., internal and kinetic) energy within a cylindrical jet section. To calculate a single number for the efficiency per model, we follow a route similar to that in the case of the deposited energy. Both internal and kinetic energies are calculated for each section of length $\lambda_{\rm max}$ at each moment in time and the results are averaged over time after the second longest-wavelength shell passes the given section. The appropriate ratio of these averages is used as the deposition efficiency for a given section and the maximum of these is used. We note that $\eta_{\rm dep}$ is not a true efficiency of the process of transferring energy from the bulk motion to the internal energy. Instead of dividing the output internal energy by the input kinetic energy we decided to use the value of the kinetic energy content co-temporal with the internal energy content. This makes $\eta_{\rm dep}$ more straightforward, but provides a different numerical value. We note that efficiency may be affected by energy composition of the fluid flowing out of the $r=3r_{\rm jet}$ boundary. In most cases, we find these outflows to be of minor importance in the energy balance of each jet section. As a thorough tracking of these outflows would require considerable numerical resources (integration over surface in a dense grid in time), we decided against including them in the definition of $\eta_{\rm dep}$.

\section{Results and Discussion}\label{sect:results}

\subsection{Qualitative description}\label{sect:qualitative}

\begin{figure*}
 \centering
 \includegraphics[resolution=600]{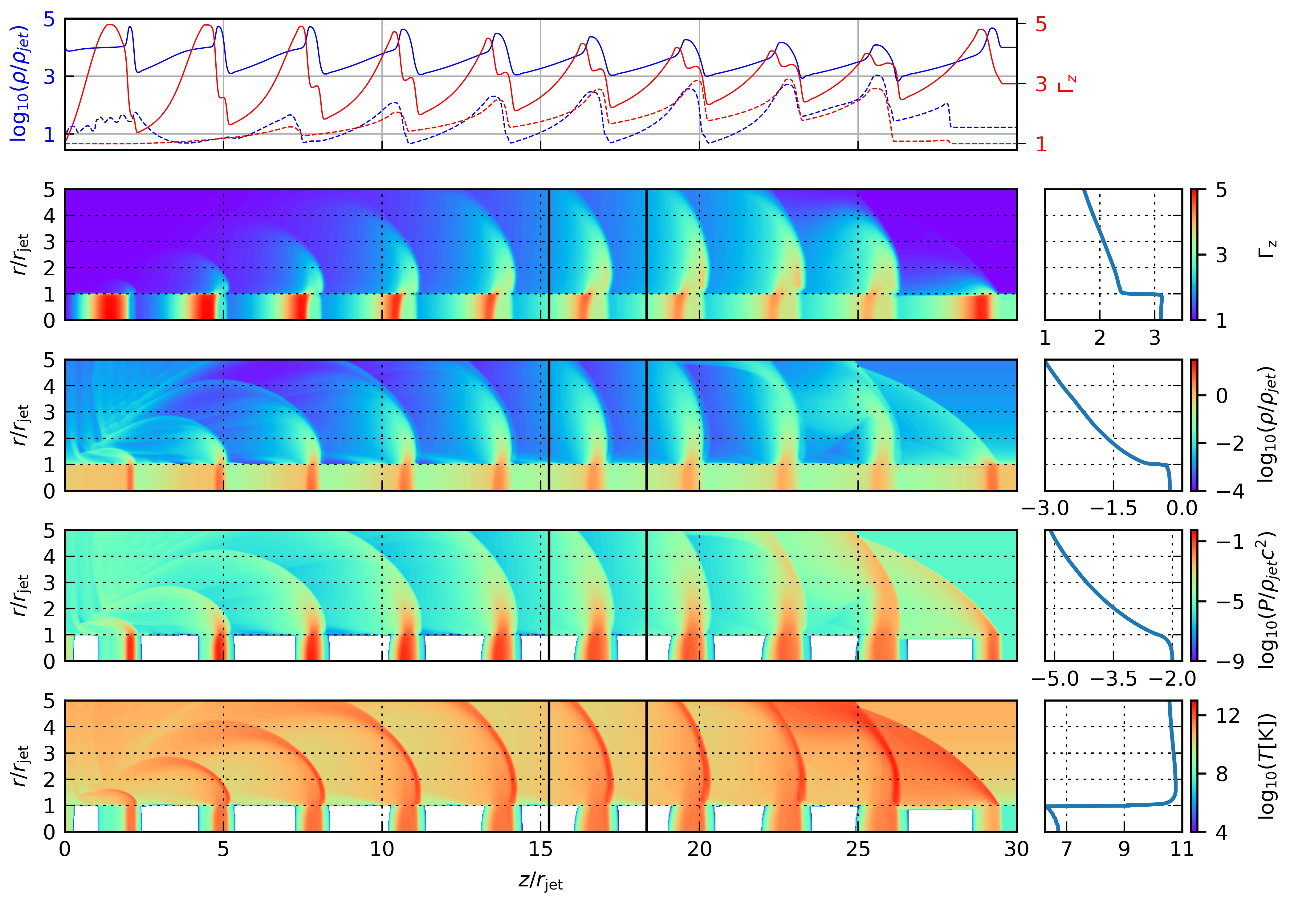}
 \caption{Propagation of a single-component pattern through the jet -- ambient medium system. The depicted case corresponds to the ``default'' case of an injection pattern with a single sinusoidal variation of the Lorentz factor between $\Gamma=1$ and $\Gamma=\Gamma_{\rm max}=5$ at the ejection site. The ejection angular frequency $\omega=2$ (in simulation units) and the initial density contrast between the jet and the ambient medium is $\rho_{\rm jet}/\rho_{\rm amb} = 10^4$. The topmost plot shows 1D slices of density (blue lines and ticks) and Lorentz factor (red lines and ticks) at the jet axis (solid lines) and at $r=3r_{\rm jet}$ (dashed lines). The remaining rows show (from top to bottom): the Lorentz factor of the motion along the jet axis $\Gamma_z$, the rest-frame density, the rest-frame pressure, and the rest-frame temperature of the fluid. In the left column, a slice of each quantity is shown at $\phi=0$. White spaces mark regions of the slice where the respective quantities fell below the color scale used -- this ``plotting floor'' was allowed in order to represent the main regions of interest in more detail. The panels in the right column show the radial profiles of each of the quantities, averaged in the $z$ direction along the wavelength of the shell ejected as fifth. The averaged area is marked on the contour plots by two black vertical solid lines. Note that the vertical axis of each profile plot shows the distance from the jet axis, matching the vertical axis of the respective contour plot. Except for the temperature profile, where averaging was performed geometrically, all the other profiles were obtained through arithmetic averages.}
 \label{fig:qualitative}
\end{figure*}

\begin{figure}
\centering
\includegraphics[resolution=600]{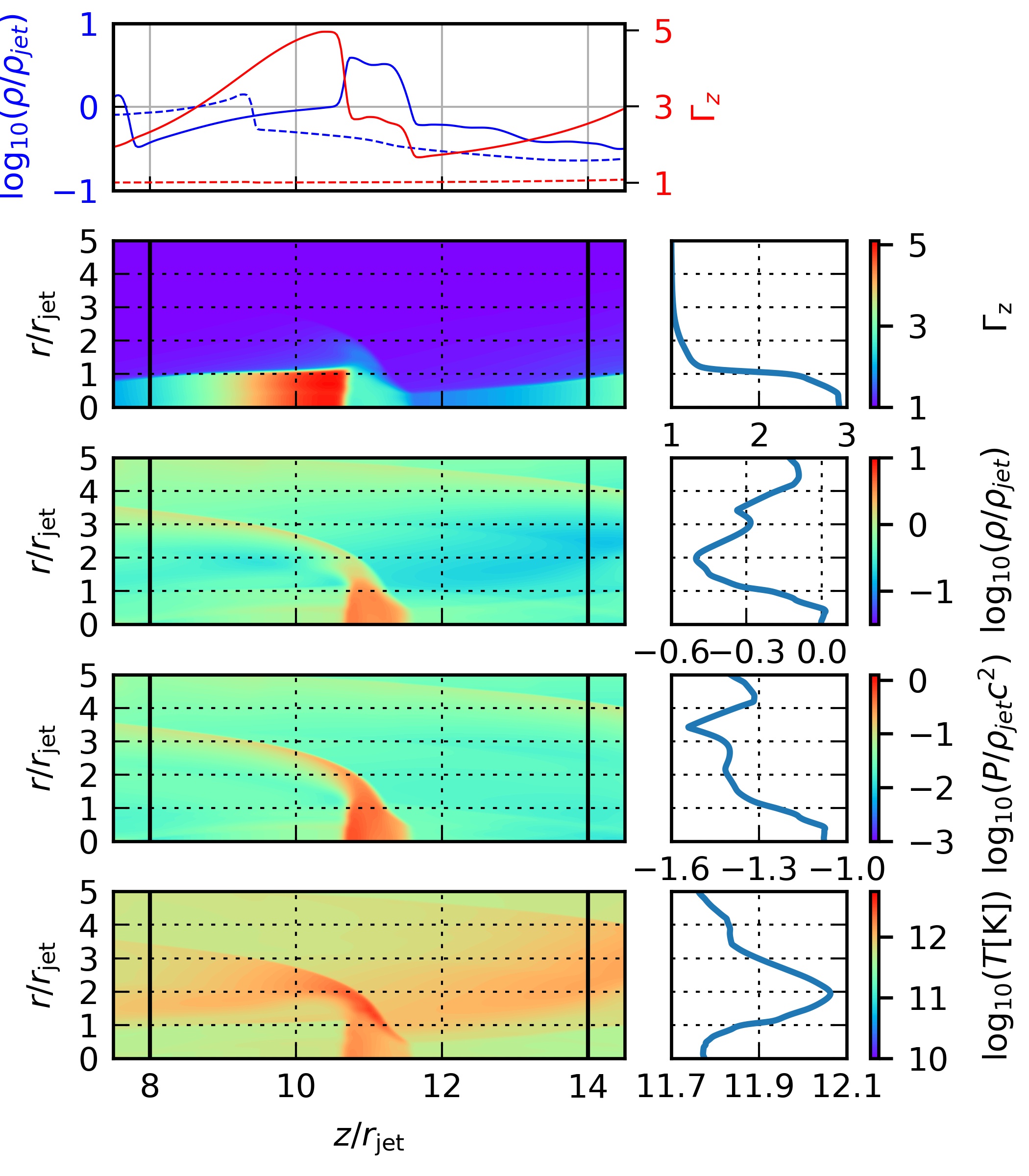}
\caption{Single shell snapshot for $\Gamma_{\rm max} = 5$, $\rho_{\rm jet}/\rho_{\rm amb} = 1$, $\omega = 1$, $t=50.5$~sim.u.~$=13.9$~ms, a case where the shock structure of a shell is well visible. See Sect.~\ref{sect:qualitative} for description. The meaning of symbols and notation are the same as in Fig.~\ref{fig:qualitative}.}
\label{fig:shellMorphology}
\end{figure}

Let us start with a general description of how the injected shells propagate through the simulation domain. The snapshots of initial and final points of the evolution of a single shell pattern are shown in Fig.~\ref{fig:qualitative}. While shells enter the simulated fluid as smooth sinusoidal variations of the Lorentz factor, they quickly steepen to form shocks within the jet. There are usually two shocks associated with each peak of the initial sinusoidal Lorentz factor variation.
Because of the non-linear dependence of velocity on the Lorentz factor, the troughs of the shell pattern will lag considerably behind, while the peaks (moving at $\sim c$) will retain their shape. The minimum of each pattern occurs by construction at $\Gamma = 1$, so a shock will form almost immediately at each minimum of the pattern. Due to relatively low velocity gradients involved, these shocks will be rather weak. However, the slower halve of the pattern evolves much faster than the velocity peaks, while enclosing the same amount of mass. Therefore, the post-shock material will quickly reach considerable density, while moving much slower than the non-shocked material of the pattern's peak (which, at that point, is either still separated from the post-shock zone merely by a contact discontinuity, or has not yet been injected into the simulation domain). The concentration of mass carries enough momentum to launch a second (reverse) shock into the fluid within the peak of the shell pattern injected after the given trough. The resulting morphology of an internal shell arising from a sinusoidal Lorentz factor variation is shown in Fig.~\ref{fig:shellMorphology}. Each shell is bounded by two shocks (forward and reverse), moving away from each other. Due to relatively large Lorentz factors involved, this shell expansion is slow in the LAB frame. The shocks are marked by regions of high temperature and trap a (common post-shock) region of high-density high-pressure fluid between them. Having no way of efficiently evacuating the shell through the shocks, this fluid heated by the shocks launches powerful streams of gas perpendicularly to the jet (henceforth: ``perpendicular outflows''). These streams retain much of the internal energy of the shocked gas over multiple jet radii, significantly extending the size of the region thermally influenced by the shocks. Moreover, the outflows launch powerful shocks into the ambient medium, which was already hot (since it was initialized as a low-density fluid of high pressure). As a result, well localized regions of very high temperatures and often significant density emerge in the ambient medium. These regions may supply conditions significantly more dramatic than the internal shocks themselves (when it comes to the generated internal energy density and Lorentz factor gradients), which may make them of interest with regard to particle acceleration and enhanced brightness (assessment of this connection is non-trivial and reaches beyond the scope of this work). While the streams are bent into bow-shocks by the interaction with the ambient medium, each consecutive bow-shock has a shielding effect on the next one (through shocking of the ambient medium into lower relative speeds). As a result, all the streams following the one ejected as second are almost identical in structure and directed at almost a right angle with respect to the jet axis, travelling downstream at the Lorentz factor of the shell that launched them (see Fig.~\ref{fig:qualitative}).

\subsubsection{Lack of the Kelvin-Helmholtz instability}

Since our model consists of two unmagnetized fluids moving at high speeds relative to each other, it seems natural to expect Kelvin-Helmholtz (KH) instabilities to occur. However, they are absent from our simulations. \cite{2007Hardee} performed a detailed analysis of the linear KH stability of jets in full MHD, and reported non-magnetized limits of his predictions. \cite{2007Mizuno} extended these considerations and confirmed the results numerically through general-relativistic MHD simulations. Since our simulations are two-dimensional, only the (axi-symmetric) pinch mode could be visible in our numerical setup. However, based on the analytical calculations of \cite{2007Hardee}, their growth rate in our entire parameter space should be negligible. On the other hand, the time scales for the low-frequency surface modes of KH instability (helical, elliptical, and higher order) can be up to four orders of magnitude shorter than the period of modulation (which is the time scale for the ejection of perpendicular outflows). These modes will grow and saturate before the modulation of the Lorentz factor can cause formation of the forward/reverse shock structure and the internal shells. However, even if these instabilities destroy the jet and maximally mix it with the ambient medium, the lateral expansion of such a mixture cannot exceed the speed of light. Meanwhile, the momentum of the structure carries it downstream and causes formation of the internal shells nonetheless, as described in our work. As a result, even a maximal KH instability of the system can only slightly broaden the jet between the internal shells, with little effect on shock formation and resulting thermalization of the bulk kinetic energy of the jet. This discussion does not apply, of course, to the laminar flow of the jet downstream from the modulation region. However, we do not expect modulation of the jet Lorentz factor to turn on as sharply as we simulate it, so this part of the jet should not be treated as physical. We also note that, in some cases, magnetic fields can stabilize the jet against the KH instability \citep[for details see][]{2007Hardee, 2007Mizuno}.

\subsection{Propagation of a single shell}\label{sect:1shell}

\begin{figure*}
 \centering
 \begin{tabular}{c|ccc}
 & $\log_{10}$(Deposited $E_{\rm int}$ & $\log_{10}$(Deposition efficiency, $\eta_{\rm dep}$) & $\log_{10}$($T_{\rm max}$ [K]) \\
 &  per wavelength [erg cm$^{-1}$]) & & \\
 \hline
 \multirow{1}{*}[12em]{\begin{sideways}$\Gamma_{\rm max} = 5$\end{sideways}} & \includegraphics{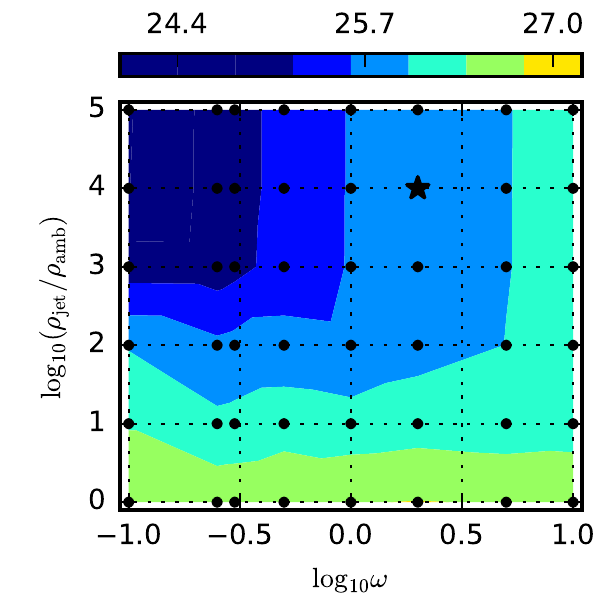} & \includegraphics{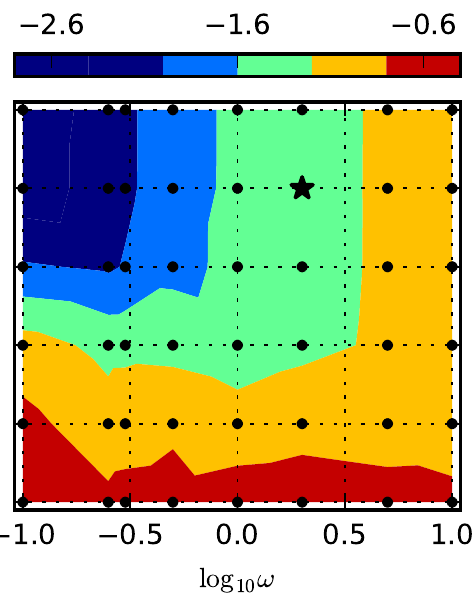} & \includegraphics{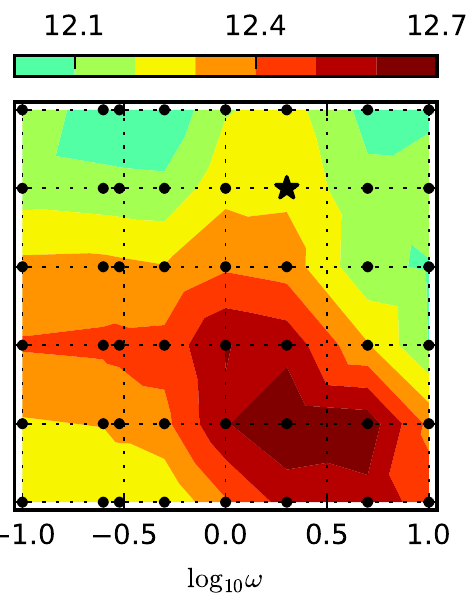} \\
 \multirow{1}{*}[12em]{\begin{sideways}$\omega = 2$\end{sideways}} & \includegraphics{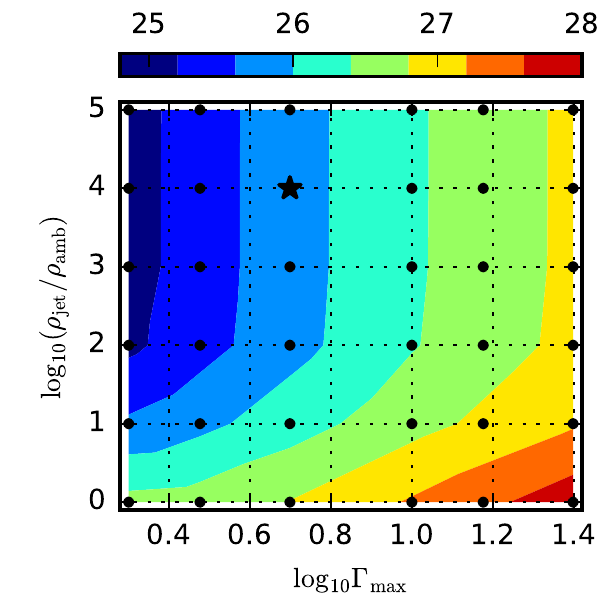} & \includegraphics{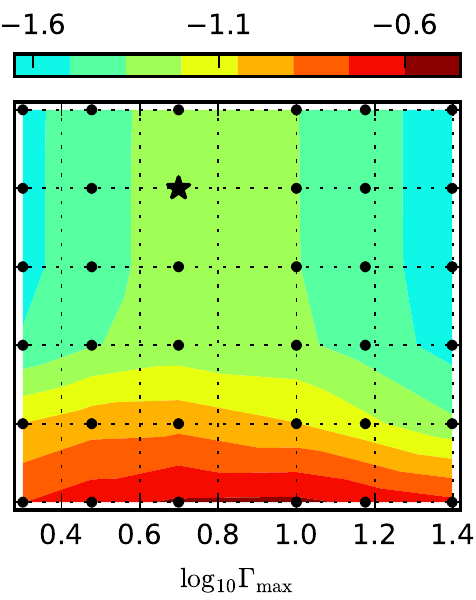} & \includegraphics{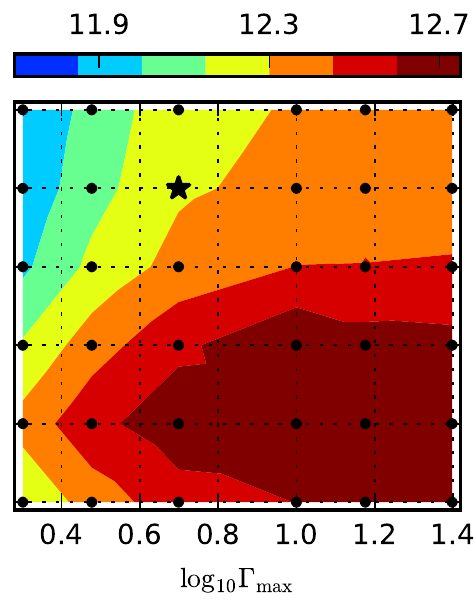} \\
 \multirow{1}{*}[12em]{\begin{sideways}$\rho_{\rm jet}/\rho_{\rm amb} = 10^4$\end{sideways}} & \includegraphics{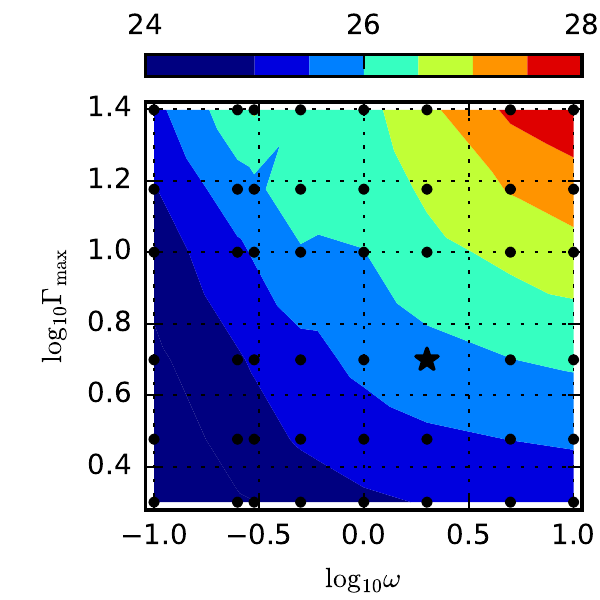} & \includegraphics{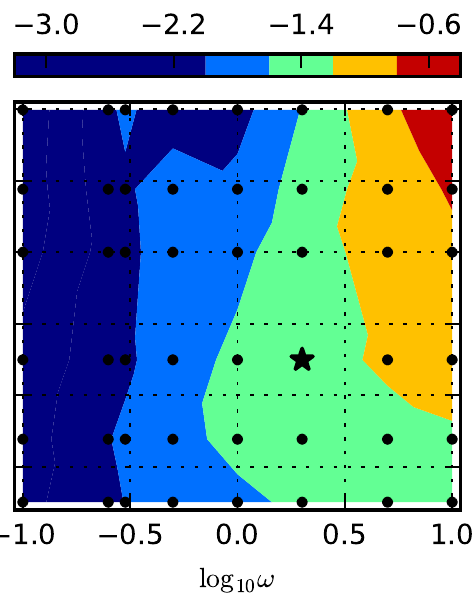} & \includegraphics{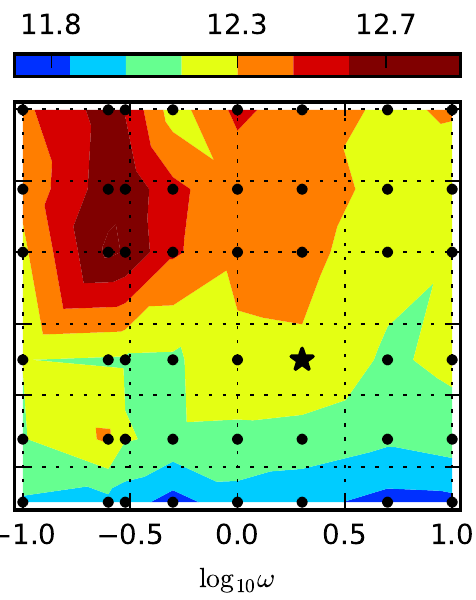}
 \end{tabular}
 \caption{Energy thermalization diagnostics. First column: contour plots of the internal energy per unit length along the jet axis deposited by a single propagating shell pattern in the LAB frame, as mapped from three slices of the parameter space. The deposited energy is averaged over time at each one-wavelength-long cylindrical section, the maximum of these averages is plotted (see Sect.~\ref{sect:diagnostics} for details). Second column: contour plots of the ratio of the LAB-frame internal energy to the total (internal and kinetic) energy in the LAB frame. Both the internal and kinetic energy are averaged over time at each one-wavelength-long cylindrical section, for each section the ratio is calculated, and the maximum of these ratios is plotted (see Sect.~\ref{sect:diagnostics} for details). Third column: the maximum temperature reached by the plasma throughout the simulation as mapped from three slices of the parameter space. The plots of each column use the same color scale. Notation: $\Gamma_{\rm max}$ -- the maximum Lorentz factor at which fluid is injected into the jet (i.e., the jet Lorentz factor at the ejection site varies between 1 and $\Gamma_{\rm max}$); $\rho_{\rm jet}$, $\rho_{\rm amb}$ -- densities of the jet and the ambient medium, respectively; $\omega$ -- (angular) frequency with which shells are being injected in units of the reciprocal of the simulation time unit ($2\pi/\omega$ is the period of the Lorentz factor modulation in sim.u.). First row: a slice with a constant maximum Lorentz factor of $\Gamma_{\rm max}=5$. Second row: a slice with a constant injection (angular) frequency of $2$ (in sim.u.). Bottom row: a slice with a constant density ratio between the jet and the ambient medium of $\rho_{\rm jet}/\rho_{\rm amb} = 10^4$. Black dots mark the parameter combinations at which simulations have been performed, the black star marks the parameter combination of the ``default'' case ($\Gamma_{\rm max} = 5$, $\rho_{\rm jet}/\rho_{\rm amb} = 10^4$, $\omega = 2$).}
 \label{fig:1shell:Eth_contour}
\end{figure*}

Let us now proceed to an analysis of how the evolution of the system depends on various parameters of the adopted variable jet model. We will proceed by considering three slices of the parameter space, with fixed $\Gamma_{\rm max} = 5$, $\rho_{\rm jet}/\rho_{\rm amb} = 10^4$, and $\omega = 2$. Note that the three slices overlap at a fiducial ``default'' model. The slices are then sampled using parameters from: $\Gamma_{\rm max} \in \{2, 3, 5, 10, 15, 25\}$, $\rho_{\rm jet}/\rho_{\rm amb} \in \{10^0, 10^1, 10^2, 10^3, 10^4, 10^5\}$, $\omega \in \{0.1, 0.25, 0.3, 0.5, 1, 2, 5, 10\}$, allowing us to identify trends in the diagnostics in relation to the model parameters.

\subsubsection{Energy deposition at a constant $\Gamma_{\rm max}$}

\begin{figure}
 \centering
 \includegraphics[resolution=600]{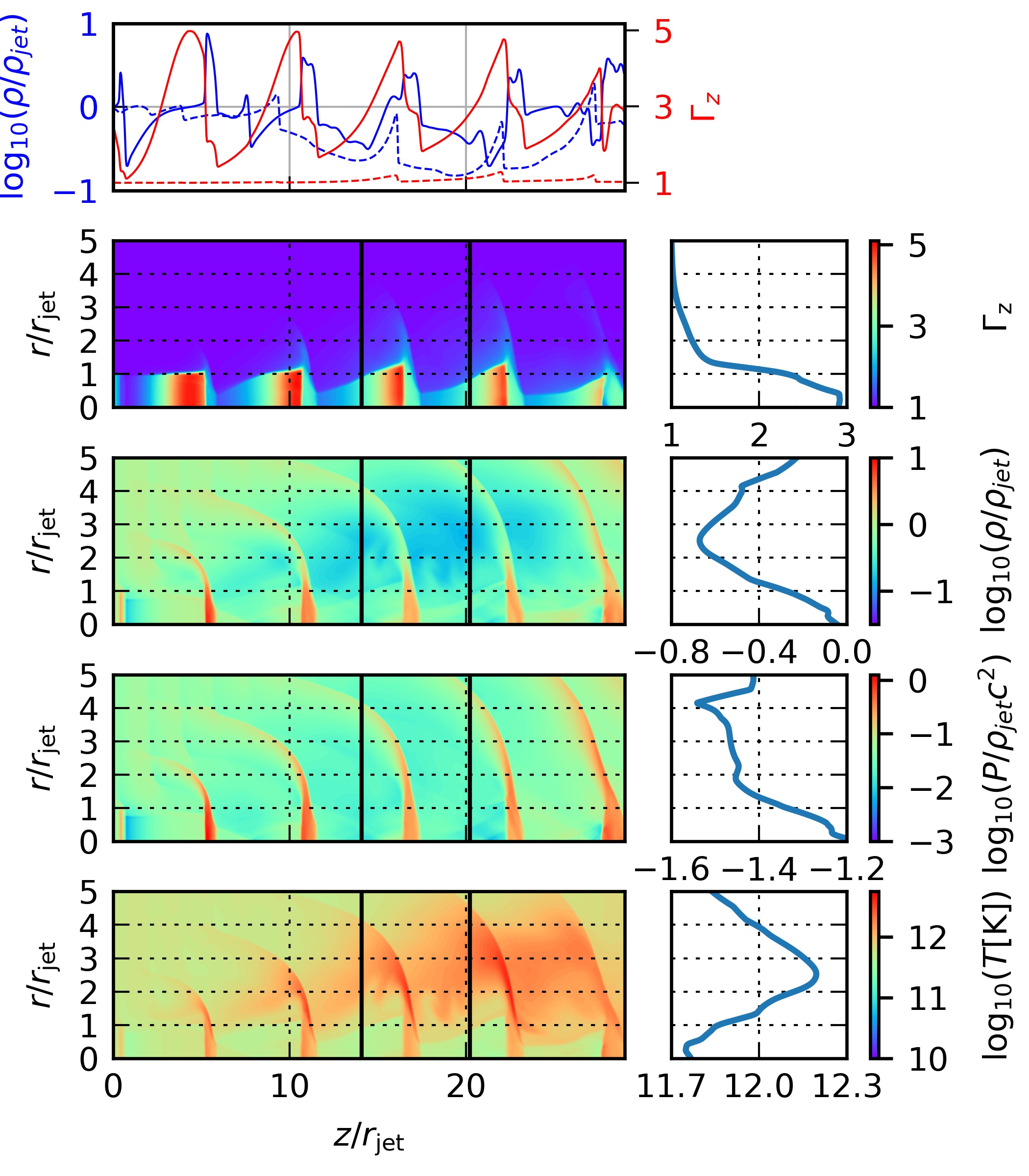}
 \caption{Shell evacuation in the single-shell model with $\Gamma_{\rm max}=5$, $\rho_{\rm jet}/\rho_{\rm amb} = 1$, $\omega = 1$, $t = 50.5$~sim.u.~$ = 13.9$~ms. The meaning of symbols and notation are the same as in Fig.~\ref{fig:qualitative}. The radial profiles are averaged over the wavelength of the shell marked with vertical black solid lines.}
 \label{fig:EdepGamConstLowRho}
\end{figure}

We start with the description of bulk motion thermalization in case of constant $\Gamma_{\rm max} = 5$, i.e., the injected Lorentz factor of the jet varying between $\Gamma = 1$ and $\Gamma = 5$. As can be seen in Fig.~\ref{fig:1shell:Eth_contour}, there is a clear boundary between the region where the density contrast does not affect the internal energy deposition at $\rho_{\rm jet}/\rho_{\rm amb} \gtrsim 50$ and the region where it is the decisive factor at lower values. At low density contrasts, the ambient medium carries significant momentum relative to the matter trapped between the shocks in each shell (see Sect.~\ref{sect:qualitative}). As a result, it is able to efficiently carry away the gas spilling out of each shell, decrease the density of the outflow, and quickly decrease the pressure in the perpendicular outflows. The hot gas is ``sucked out'' of the shells, slowed down by the shocks between the outflows and ambient medium, and finally well mixed with the ambient medium. This accounts for efficient thermalization of the jet's bulk kinetic energy, as seen in Fig.~\ref{fig:EdepGamConstLowRho} for the case of $\rho_{\rm jet}/\rho_{\rm amb} = 1$, $\omega = 1$. Since the final well-mixed fluid quickly loses information about its initial structure, the energy deposition at low density contrast is also effectively independent of the Lorentz factor modulation frequency (Fig.~\ref{fig:1shell:Eth_contour}).

\begin{figure}
 \centering
 \includegraphics[resolution=600]{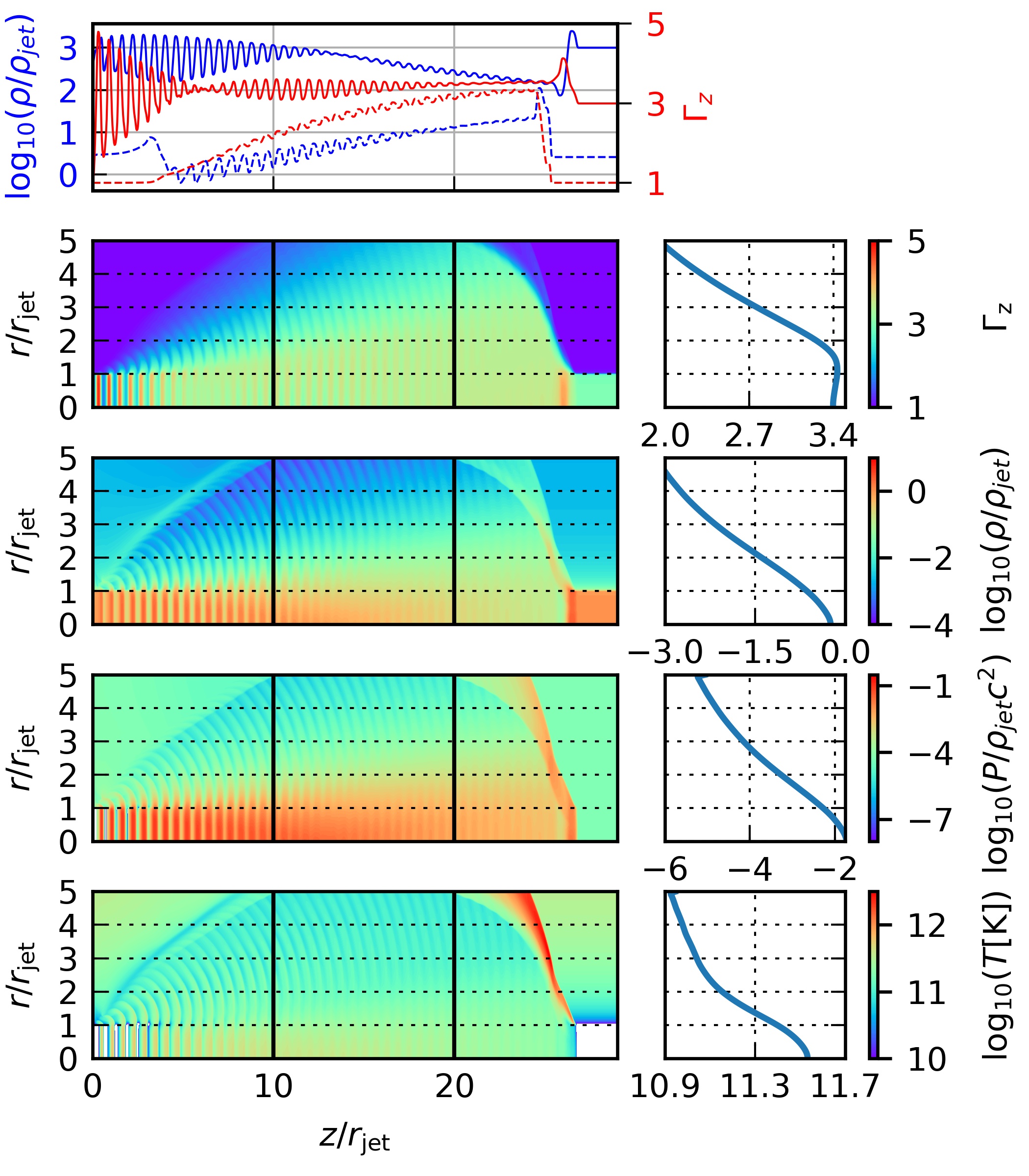}
 \caption{Thermalization in the single-shell model with $\Gamma_{\rm max}=5$, $\rho_{\rm jet}/\rho_{\rm amb} = 10^3$, $\omega = 10$, $t = 27.5$~sim.u.~$ = 7.6$~ms. The meaning of symbols and notation are the same as in Fig.~\ref{fig:qualitative}. Radial profiles are averaged over $z/r_{\rm jet} \in [10, 20]$ (as marked by the two black vertical solid lines).}
 \label{fig:EdepGamConstHighNu}
\end{figure}

At a density contrast of $\rho_{\rm jet}/\rho_{\rm amb} \gtrsim 50$, increasing the modulation frequency leads to increased internal energy generation, by about a factor of $3$ between $\omega = 0.1$ and $\omega = 10$, as well as increased thermalization efficiency. This trend can be again attributed to more efficient mixing of the fluid, which occurs, however, by means of a different mechanism. At low ejection frequencies, powerful shocks formed between the perpendicular outflows and the ambient medium are able to increase pressure in the outflows and prevent efficient evacuation of the shells (when combined with the fact that the low density of the ambient medium does not allow for the outflows to be efficiently carried away from the jet). This results in inefficient mixing and thermalization. At high injection frequencies, the shielding effect the outflows have on each other (see Sect.~\ref{sect:qualitative}) prevents efficient shocks in the ambient medium from forming, and allows the outflows to freely disrupt the jet. Due to their close proximity, the outflows then shock against each other. While these shocks do not seem to be important dynamically, they provide a means of efficient and near-homogeneous thermalization of the kinetic energy carried by the outflows. An extreme example of this effect is shown for $\rho_{\rm jet}/\rho_{\rm amb} = 10^3$, $\omega = 10$ in Fig.~\ref{fig:EdepGamConstHighNu}.

\subsubsection{Energy deposition at a constant injection frequency}\label{sect:res:single:nuConst}

\begin{figure}
 \centering
 \includegraphics[resolution=600]{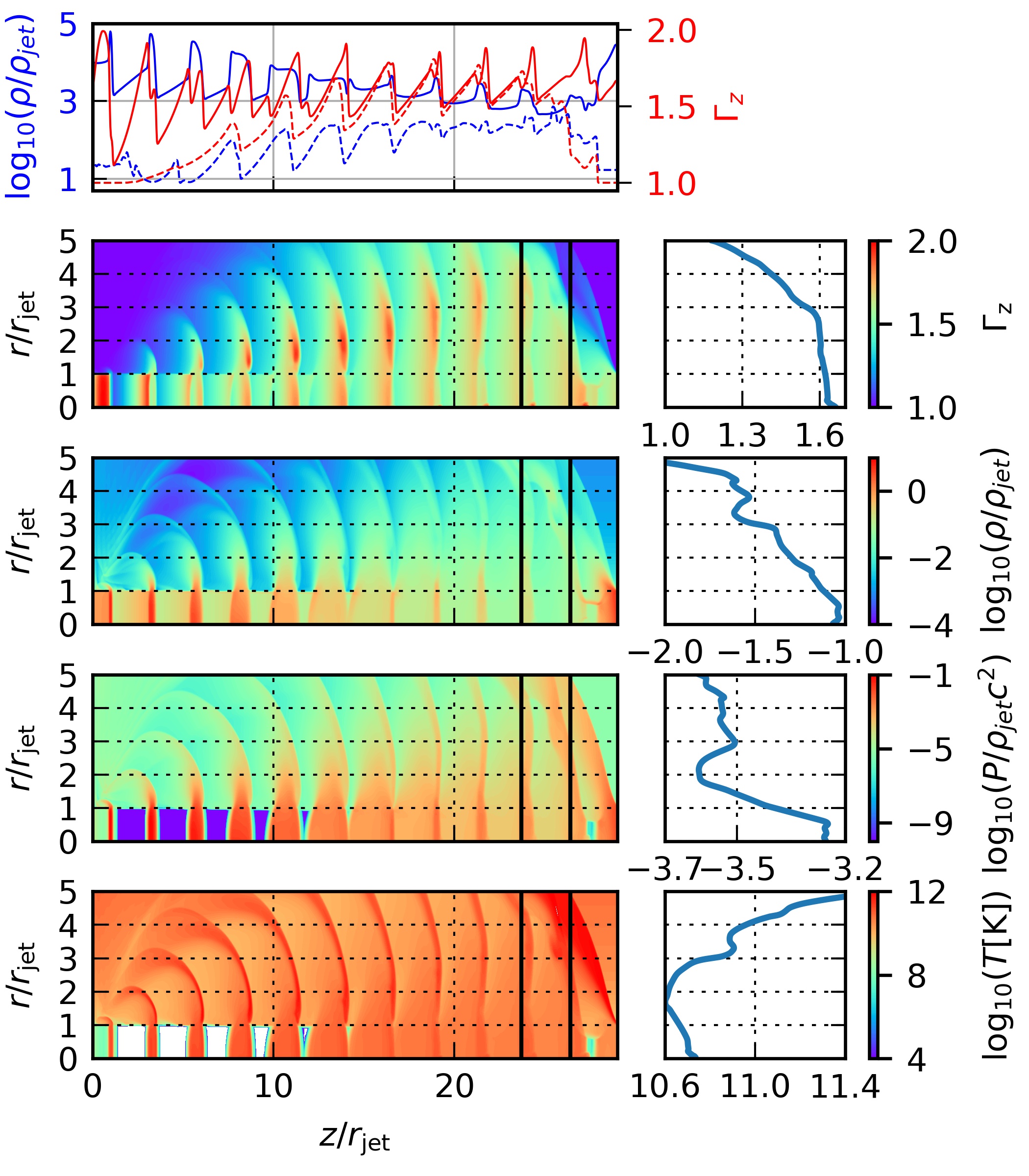}
 \caption{A snapshot showing the perpendicular outflows at low Lorentz factors, for $\Gamma_{\rm max}=2$, $\rho_{\rm jet}/\rho_{\rm amb} = 10^4$, $\omega = 2$, $t=36.0$~sim.u.~$ = 9.9$~ms. The meaning of symbols and notation are the same as in Fig.~\ref{fig:qualitative}.}
 \label{fig:EdepNuConstLowGam}
\end{figure}

\begin{figure}
 \centering
 \includegraphics[resolution=600]{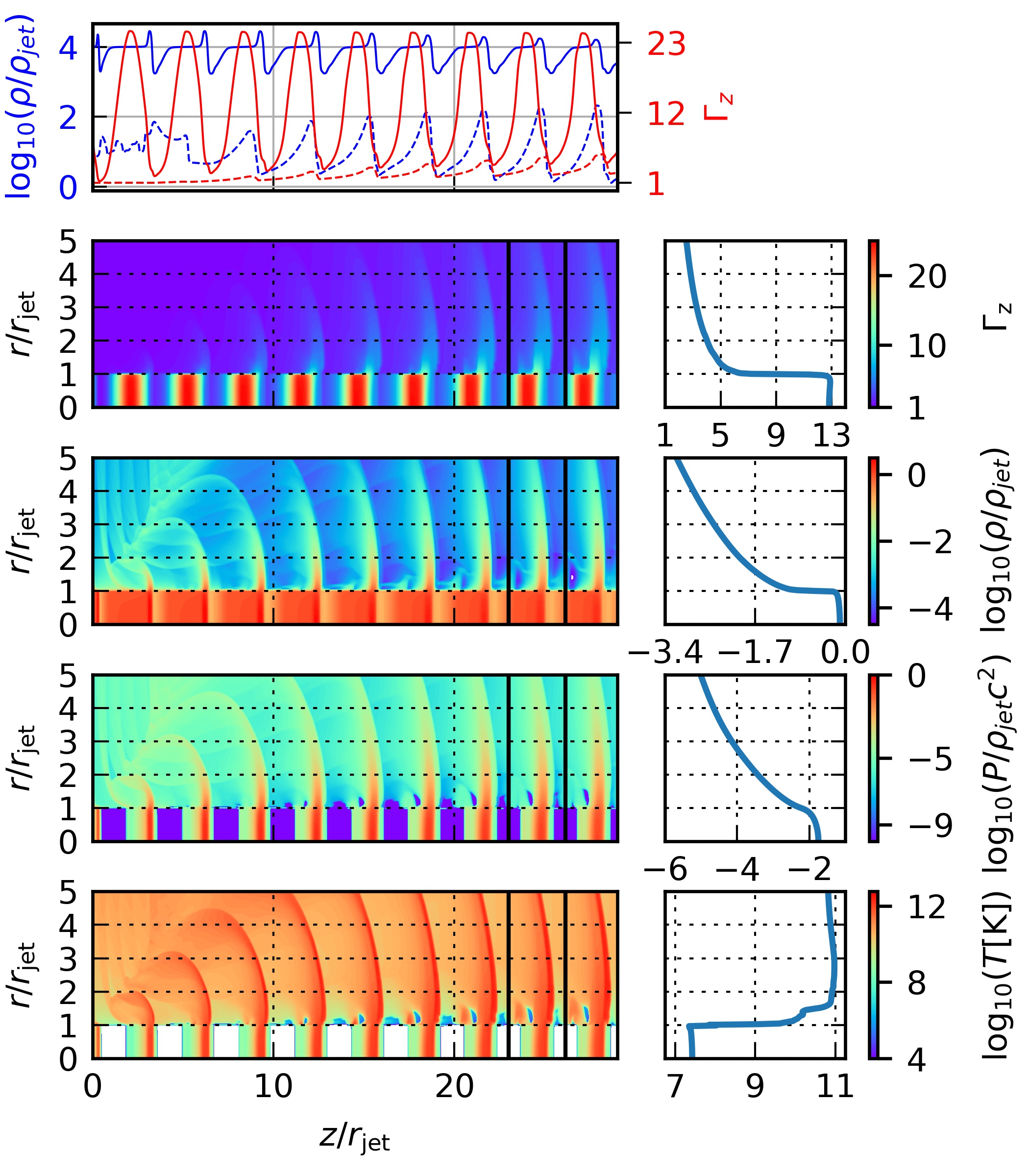}
 \caption{A snapshot showing the perpendicular outflows at high Lorentz factors, for $\Gamma_{\rm max}=25$, $\rho_{\rm jet}/\rho_{\rm amb} = 10^4$, $\omega = 2$, $t=50.5$~sim.u.~$ = 13.9$~ms. The meaning of symbols and notation are the same as in Fig.~\ref{fig:qualitative}.}
 \label{fig:EdepNuConstHighGam}
\end{figure}

We continue our analysis with the results for a fixed injection frequency $\omega=2$ (middle row of Fig.~\ref{fig:1shell:Eth_contour}). The amount of the deposited energy again decreases with increasing density contrast. Interestingly, the transition from the deposited energy being almost independent of density contrast to the sensitivity regime occurs again at $\rho_{\rm jet} / \rho_{\rm amb} \sim 50$. This suggests that this value holds a more general meaning. The deposited internal energy increases with maximal Lorentz factor as a power law for both the density-independent and density-dependent regimes.
At each $\rho_{\rm jet}/\rho_{\rm amb}$ there is a maximum in the efficiency of thermalization at $\Gamma_{\rm max} \sim 6$. Let us consider its origins. At high $\Gamma_{\rm max}$, the peak of the shell pattern carries an enormous amount of momentum, even in comparison with that of the dense post-shock material of the forward shock of the shell (see Sect.~\ref{sect:qualitative}). As a result, the reverse shock is never launched and the shell material is not trapped between the shocks. While the perpendicular outflows still form (Fig.~\ref{fig:EdepNuConstHighGam}), they are mostly composed of material moving slowly with respect to the bulk of the jet and of relatively low density and pressure. These outflows do generate shocks within the ambient medium, but the amount of the internal energy generated is well below the bulk kinetic energy of the jet, still mostly enclosed within $r<r_{\rm jet}$ due to the low efficiency of perpendicular outflows in evacuating the jet. At low maximum Lorentz factors, the forward and reverse shocks move away from each other at noticeable LAB-frame velocities (due to the low speed of the flow, their speed is not close enough to $c$) and quickly decompress the post-shock region (Fig.~\ref{fig:EdepNuConstLowGam}). The perpendicular outflows, while initially fast, evolve into broad features providing significant shielding to each other (see Sect.~\ref{sect:qualitative}). This prevents powerful shocks from forming in the ambient medium, which is mixed with the jet matter by the broad outflows without efficient thermalization of the kinetic energy.

\subsubsection{Energy deposition at a constant density contrast}

Let us now consider the third slice through our parameter space, at a fixed density contrast of $\rho_{\rm jet}/\rho_{\rm amb} = 10^4$ (the bottom row of Fig.~\ref{fig:1shell:Eth_contour}). Note that the density contrast is well above the limit of $\rho_{\rm jet} / \rho_{\rm amb} \sim 50$, so the internal energy deposition should not be considerably dependent on the density contrast at this point. We see that the amount of deposited internal energy increases much steeper with the Lorentz factor than with injection frequency. This effect appears to be caused mainly by the reservoir of available kinetic energy increasing with the Lorentz factor, as in the case of fixed-$\omega$ slice. Once the influence of change in the available total energy is removed by calculating $\eta_{\rm dep}$, the dependence on $\Gamma_{\rm max}$ becomes much weaker. The thermalization efficiency dependence on $\Gamma_{\rm max}$ has a maximum, with its position increasing from $\Gamma_{\rm max} \sim 4$ at $\omega \sim 0.1$ to $\Gamma_{\rm max} > 25$ at $\omega=10$. This is the same maximum as the one in the case of fixed $\omega = 2$ discussed previously. As shown in the previous section, the position of this maximum is dictated by the jet's ability to maintain a well-defined shell morphology with forward and reverse shocks. The inability of shell patterns at high $\Gamma_{\rm max}$ to form reverse shocks should not significantly depend on injection frequency -- the relative momenta of the peak- and trough-portion of the shell pattern scale in the same way with changing wavelength. On the other hand, dissipation of the internal shells in the case of low Lorentz factors will occur faster if the shells are closer together (the shock speeds at a given $\Gamma_{\rm max}$ should be comparable). Consequently, the thermalization at a higher $\omega$ in low-$\Gamma_{\rm max}$ case will be shut off sooner than at low-frequencies. This dependence causes the maximum internal energy deposition efficiency to move towards higher $\Gamma_{\rm max}$ with increasing frequency.

When it comes to the dependence on $\omega$, it is quite clear that thermalization efficiency within $3r_{\rm jet}$ increases with ejection frequency and is very sensitive to this parameter, especially at high Lorentz factors and high $\omega$. This is likely related to the amount of power available to perpendicular outflows. At low injection frequencies, the buildup of mass in the internal shell is slow (it takes more time for a peak of the shell pattern to catch up with the points of lowest velocity) and the outflows have plenty of time to lower pressure between the shocks before additional matter gathers there. As a result, the outflows are ejected by small pressure gradients and are not able to drive strong shocks into the ambient medium which would turn the kinetic energy they carry into internal energy. Growing $\Gamma_{\rm max}$, in addition to preventing dissipation of internal shocks inside the jet (see Sect.~\ref{sect:res:single:nuConst}), increases the efficiency with which bulk kinetic energy is turned into internal energy in the shell zone between the forward and reverse shocks. This, in turn, increases the pressure gradient available to the perpendicular outflows and their ability to drive shocks into the ambient medium.

\subsubsection{Maximal temperature at a constant $\Gamma_{\rm max}$}

The contour plots summarizing maximal temperatures found in our simulations are shown in the rightmost column of Fig.~\ref{fig:1shell:Eth_contour}. The first notable result is the fact that maximum temperatures remain within an order of magnitude of $\sim 10^{11.9}$~K for all the tested parameter space. As the ambient medium was initialized at the same temperature for all simulations, $T_{\rm amb, init} = P_{\rm amb}\mu m_p / (\rho_{\rm amb}k) \simeq 10^{11.78}$ (cf., Sect.~\ref{sect:numericalSetup:initialization}), this is not unexpected. It appears that while the shocks driven by perpendicular outflows into the ambient medium can deposit considerable amounts of internal energy, they cannot significantly increase the temperature of the, already hot, gas surrounding the jet. We note that relativistic jets are expected to be surrounded by such hot gas in form of cocoons formed during propagation through the interstellar medium \citep[e.g., ][]{1989Begelman}, so, to some extent, this observation should be true for real astrophysical environments.

\begin{figure}
 \centering
 \includegraphics[resolution=600]{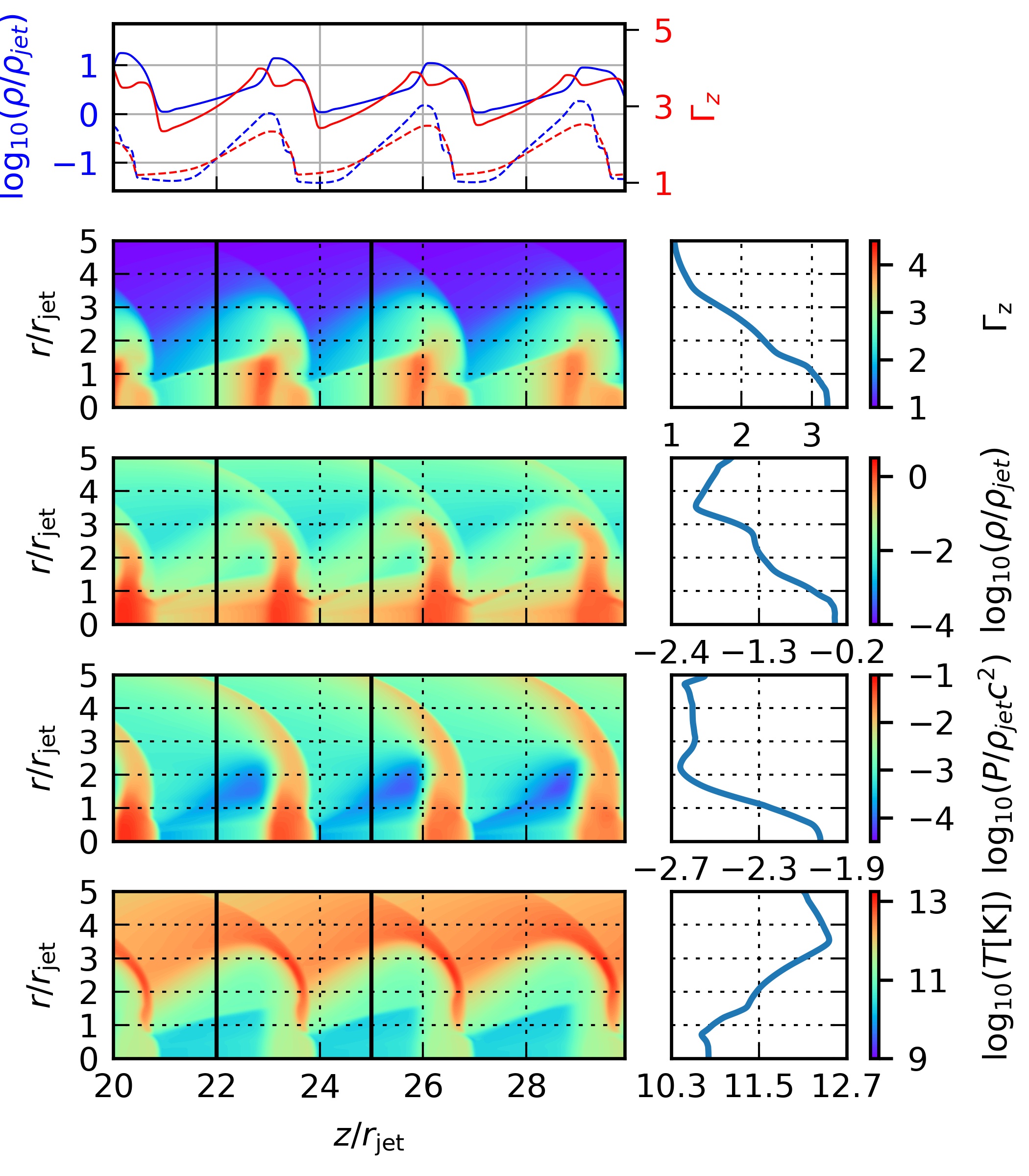}
 \caption{A snapshot of the maximal temperature maximum, for $\Gamma_{\rm max}=5$, $\rho_{\rm jet}/\rho_{\rm amb} = 10$, $\omega = 2$, $t=50.5$~sim.u. The meaning of symbols and notation are the same as in Fig.~\ref{fig:qualitative}.}
 \label{fig:TmaxGamConstMax}
\end{figure}

Once we limit our considerations to a slice with $\Gamma_{\rm max} = 5$, we find that there is a clear maximum of maximal temperature at $\omega \sim 2$, $\rho_{\rm jet} / \rho_{\rm amb} \sim 10$. This maximum is related to the emergence of a well defined shock structure driven by the perpendicular outflows, efficiently compressing the ambient medium to jet-like densities. At these values of model parameters, such structures are atypically dense, but still moving downstream at a high Lorentz factor. A snapshot of the model at the position of this maximum is shown in Fig.~\ref{fig:TmaxGamConstMax}. At frequencies below the maximum, the hydrodynamical properties at the edges of perpendicular outflows are continuous due to large wavelengths of the Lorentz factor modulation and motion of the shock fronts unloading matter from the internal shells. As a result, only weak shocks are driven into the ambient medium (Fig.~\ref{fig:TmaxRhoConstLowNu}). At high frequencies, the gaseous streams are spaced closely together and merge with each other, preventing strong shocks from appearing as well (see Fig.~\ref{fig:EdepGamConstHighNu}). At low density contrasts, the jet is destroyed and mixed with the ambient medium before notable amounts of gas can gather in the internal shells (Fig.~\ref{fig:EdepGamConstLowRho}). Finally, at high density contrasts, the ambient medium quickly becomes a mix of jet gas and post-shock material, which causes pre-shock conditions for each consequent shell to be less favorable for a high post-shock temperature (see Fig.~\ref{fig:qualitative}).

\subsubsection{Maximal temperature at constant injection frequency}

As seen in the second row of Fig.~\ref{fig:1shell:Eth_contour}, at any Lorentz factor the maximum temperature reaches a maximum at a density contrast of $\sim 10$. This is the same maximum as the one observed in the $\Gamma_{\rm max} = 5$ slice and the reasons for its dependence on the density contrast have been discussed above.

When it comes to the dependence on the maximum Lorentz factor, the temperature's behavior seems to be divided into two regimes. At $\Gamma_{\rm max} \lesssim 10$, the maximum temperature increases with the Lorentz factor modulation amplitude by a factor of $\sim 3$. This is caused by the ``relative speed'' of each shell's forward and reverse shocks decreasing in the LAB frame (i.e., both of them asymptotically approaching $c$). This improves the confinement of the high-pressure gas in each shell, which leads to more powerful outflows launching stronger shocks into the ambient medium. At $\Gamma_{\rm max} \gtrsim 10$, both forward and reverse shocks are essentially moving at $c$ in the LAB frame and the maximum reached temperature is independent of the modulation amplitude.

It is interesting to note that the maximum temperature remains sensitive to the density contrast up to very high relative densities of $\rho_{\rm jet}/\rho_{\rm amb} \sim 10^3$. The maximum temperature in our models is always reached in the shocks launched into the ambient medium by the gaseous streams ejected perpendicularly to the jet axis. It is therefore not surprising that properties of the ambient medium influence the maximum temperature even at high density contrasts. Eventually, when the ambient medium density is low enough, the region of maximum temperature should move to the shocks inside the jet, at which point the dependence of $T_{\rm max}$ on $\Gamma_{\rm max}$ should change sharply.

\subsubsection{Maximal temperature at a constant density contrast}

\begin{figure}
 \centering
 \includegraphics[resolution=600]{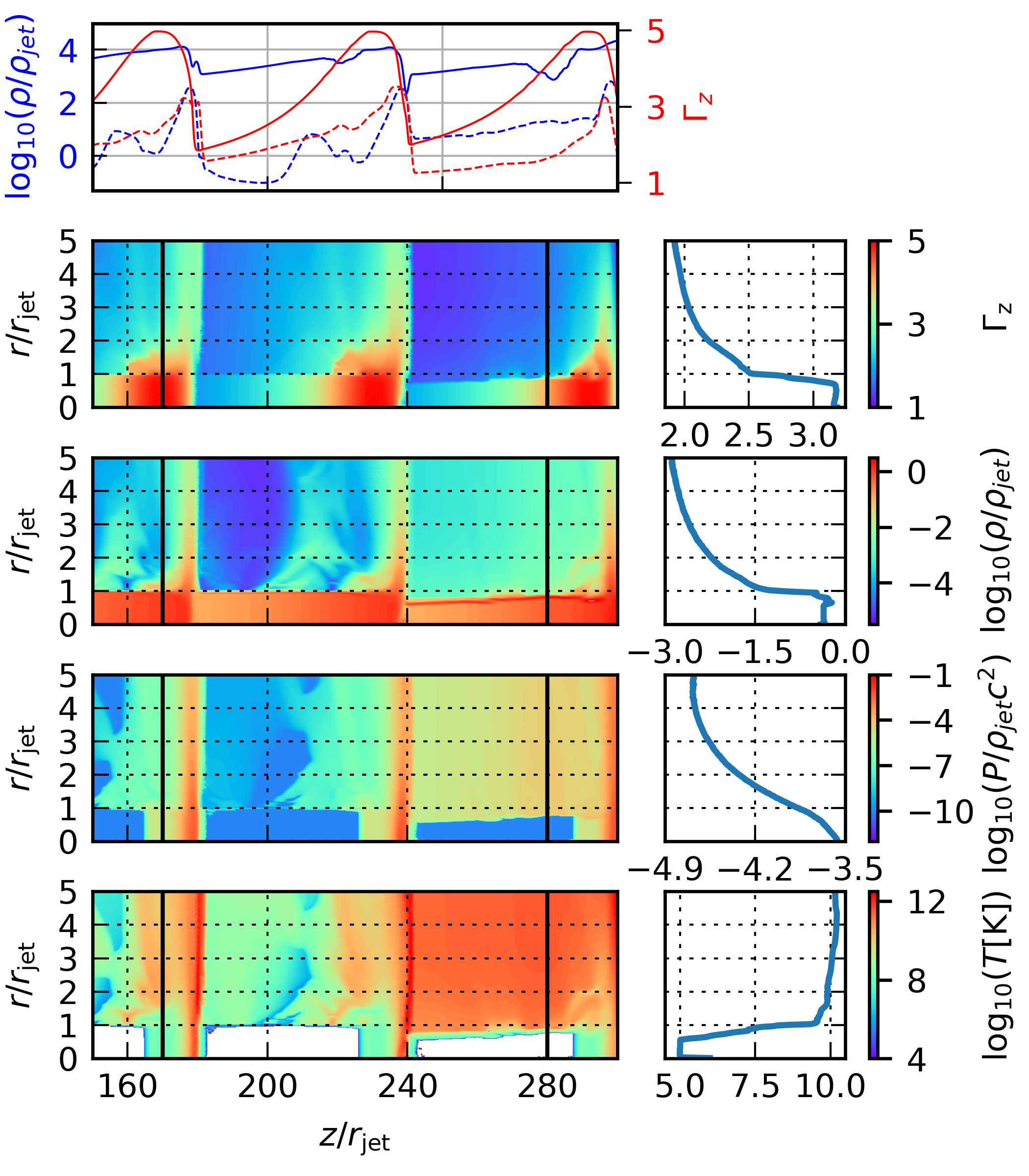}
 \caption{A snapshot of the maximal temperature for $\Gamma_{\rm max}=5$, $\rho_{\rm jet}/\rho_{\rm amb} = 10^4$, $\omega = 0.1$, $t=317.3$~sim.u. The meaning of symbols and notation are the same as in Fig.~\ref{fig:qualitative}.}
 \label{fig:TmaxRhoConstLowNu}
\end{figure}

The rightmost bottom plot of Fig.~\ref{fig:1shell:Eth_contour} shows the maximum temperature of the fluid for the slice at a constant density contrast of $10^4$. As in the case of a constant injection frequency, we see that the maximal temperature increases rapidly with $\Gamma_{\rm max}$ for $\Gamma_{\rm max} \lesssim 10$ and becomes almost independent of the Lorentz factor modulation amplitude at higher values. It is interesting to note that the rise with $\Gamma_{\rm max}$ is steepest at $\omega = 0.25$, in accordance with the fact that, at each $\Gamma_{\rm max}$, the maximum temperature reaches a maximum at $\omega = 0.25$ (at $\Gamma_{\rm max} = 2$ the maximum temperature does not depend on the ejection frequency since the jet is destroyed before the ambient medium shocks can be created). At low frequencies (see Fig.~\ref{fig:TmaxRhoConstLowNu}), this is likely due to the long time needed for the internal shocks to steepen, which causes the gaseous outflows to be broad, low-density structures, not able to drive strong shocks into the ambient medium. The nature of the high-frequency part of this dependence is likely the same as in the case of the maximum of $T_{\rm max}$ in the $\Gamma_{\rm max} = 5$ slice. There, the mixing occurs due to the gaseous streams launched from the internal shocks (see Fig.~\ref{fig:EdepGamConstHighNu}) preventing well-defined ambient medium shocks from forming through shielding effects (see Fig.~\ref{fig:EdepGamConstHighNu}, Sect.~\ref{sect:qualitative}).

\subsubsection{Mach numbers of the shocks}

As the system evolves, the material between internal shells is quickly decompressed due to its initial motion (see, e.g., Fig.~\ref{fig:qualitative}). It is often also relatively primordial, not heated by the shocks, and so its sound speed is low compared to the speed of light. At the same time, the internal shocks can be considerably relativistic and their speed, even relative to the moving inter-shell gas, remains close to $c$.

We have calculated Mach numbers of the forward and reverse shocks within each of our simulations. For forward shocks, the Mach number remains stable to within an order of magnitude for each simulation run and reaches values $10^4$-$10^5$ for most of our parameter space. The Mach numbers are lower for high frequencies and low density contrasts -- there, the forward shock may be weak, or even not form at all, as discussed in previous sections. The forward shock Mach number seems relatively insensitive to the amplitude of initial Lorentz factor modulation of the flow.

While the Mach numbers of the reverse shocks start at values comparable to those of the forward shocks, $10^4$-$10^5$, they decay by about an order of magnitude by the end of each simulation. As in the case of forward shocks, they are low for high modulation frequencies and low density contrasts. In addition, however, they exhibit a dependence on the Lorentz factor modulation amplitude. For low $\Gamma_{\rm max}$, reverse shock Mach numbers are lower, and the shocks may even not form. This has been discussed in previous sections as the inability of the system to form well-defined forward-reverse shock structures at low modulation amplitudes.

\subsection{Two-component patterns}\label{sect:2shell}

\begin{figure*}
 \centering
 \begin{tabular}{cccc}
 $\log_{10}$(Steady-state deposited $E_{\rm int}$ & $\log_{10}$(Deposition efficiency, $\eta_{\rm dep}$) & $\log_{10}$($T_{\rm max}$ [K]) \\
 per wavelength [erg/cm]) & & \\
 \hline
 \includegraphics{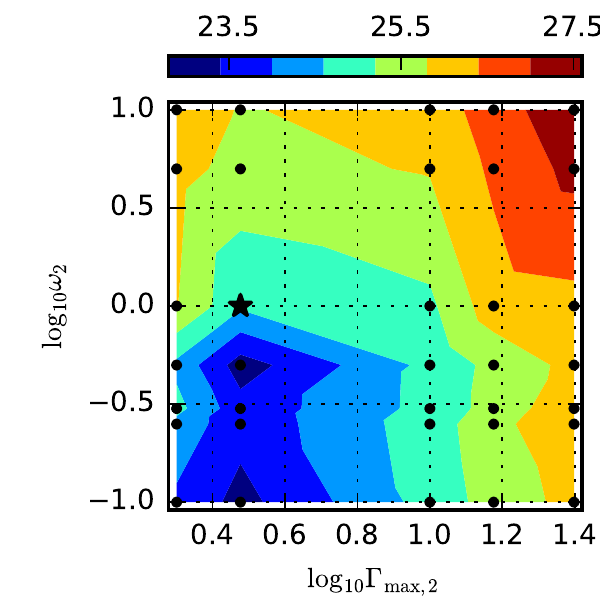} & \includegraphics{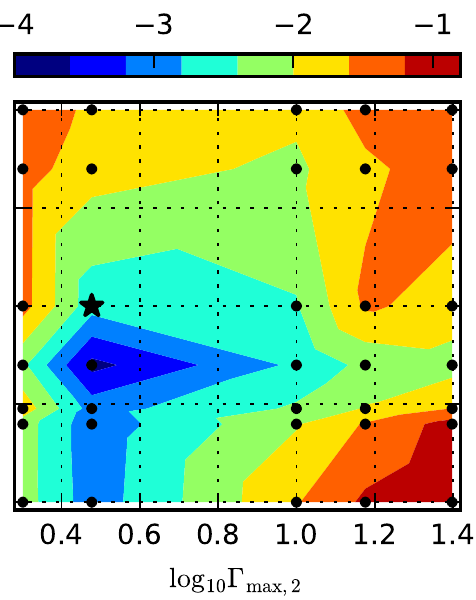} & \includegraphics{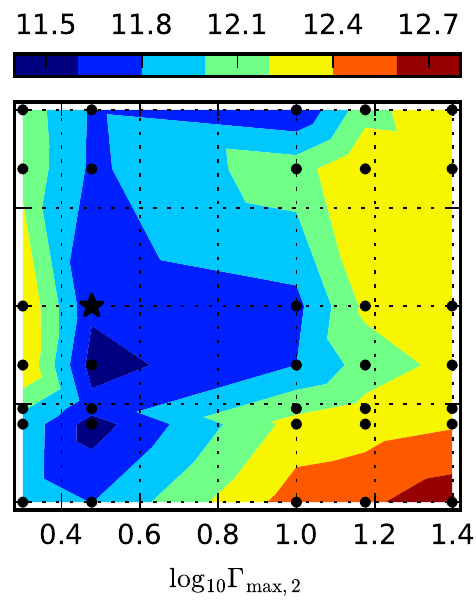} \\
 \end{tabular}
 \caption{Energy thermalization diagnostics for the two-shell collision models. The maximum Lorentz factor of the first component is fixed at $\Gamma_{\rm max, 1} = 5$, while $\Gamma_{\rm max, 2}$ is allowed to vary. Similarly, injection frequency of the first modulation component is fixed in all cases at $\omega_1 = 2$. Injection frequency of the second modulation component, $\omega_2$, is allowed to vary. For all the simulations, the density contrast is equal to $\rho_{\rm jet}/\rho_{\rm amb} = 10^4$. First column: contour plots of the internal energy per unit length along the jet. The deposited energy is averaged over time at each cylindrical section of length equal to the wavelength of the lower-frequency modulation component, the maximum of these averages is plotted (see Sect.~\ref{sect:diagnostics} for details). Second column: contour plots of the ratio of the LAB-frame internal energy to the total energy (internal and kinetic) in the LAB frame. Both the internal and kinetic energy are averaged over time at each one-wavelength-long cylindrical section (in the wavelength of the low-frequency modulation component). For each section the ratio is calculated, and the maximum of these ratios is plotted (see Sect.~\ref{sect:diagnostics} for details). Third column: maximum temperature reached during the simulation. Black dots mark the parameter combinations at which simulations have been performed.}
 \label{fig:2shell:Eth_contour}
\end{figure*}

\subsubsection{Deposited internal energy as a function of the relative speed of the pattern components}

We now consider models with Lorentz factor variation composed of two sinusoidal components. Here, the peak Lorentz factor of the first component is fixed at $\Gamma_{\rm max, 1} = 5$ and its frequency at $\omega_1 = 2$, while the peak Lorentz factor of the second, $\Gamma_{\rm max, 2}$, and its frequency, $\omega_2$, are allowed to vary.

As can be seen in Fig.~\ref{fig:2shell:Eth_contour}, both the deposited energy and the deposition frequency have a minimum close to where the Lorentz factors of the two components would be the same (see also Figs.~\ref{fig:Two-EdepExtNu}, \ref{fig:Two-EdepExtNu2}, and~\ref{fig:Two-EdepExtGam} for snapshots of simulation runs surrounding the minimum). This suggests that interaction between the shells are an important factor in thermalization of the bulk kinetic energy. The minimum is most pronounced for low frequencies of the second component, i.e., thermalization is most sensitive to the Lorentz factor of the second component when this component is the long-wavelength part of the modulation. In this regime, the flow will be composed of a fringe of well-defined shells formed from the short-wavelength pattern $(\omega_1, \Gamma_{\rm max, 1})$, whose maximal Lorentz factor is modulated by $\Gamma_{\rm max, 2}$ (see, e.g., the bottom panel of Fig.~\ref{fig:Two-EdepExtNu}). Due to this secondary modulation, the shells can catch up to each other and collide, forming secondary internal shocks that aid thermalization. The efficiency with which these secondary shocks transform their bulk kinetic energy into internal energy will depend on differences in their Lorentz factors. These, in turn, are set by their modulation with the Lorentz factor of the long-wavelength pattern component, which causes sensitivity of our thermalization diagnostics to this parameter.

Within the tested region of the parameter space, the thermalization diagnostics do not appear to asymptote to values dominated by one component of the pattern. Instead, at each point the deposited internal energy and the deposition efficiency seem to be dependent on parameters of both the injected components.

The internal energy deposition efficiency reaches a minimum at $\omega_2 \sim 0.5$ for $\Gamma_{\rm max, 2} \sim 5$, corresponding to a minimum in the dependence of the internal energy deposition on $\Gamma_{\rm max, 2}$. Simulation snapshots for the low-frequency ($\omega_2 = 0.1$), minimal internal energy deposition frequency ($\omega_2 = 0.3$), and high-frequency ($\omega_2 = 1$) cases with $\Gamma_{\rm max, 2} = 10$ are shown in Figs.~\ref{fig:Two-EdepExtNu} and~\ref{fig:Two-EdepExtNu2} (we choose this value of the Lorentz factor to best visualize frequency dependence). At low $\omega_2$, below the minimum, the jet is quickly disrupted by extremely powerful and fast perpendicular outflows, directing most of the jet density far from the jet axis. 
The fringe of low-amplitude high-frequency shocks circumvents the constraint preventing the $\Gamma_{\rm max}=10$ outflows from forming well defined shells (see Sect.~\ref{sect:1shell}), which allows these powerful outflows to form. The short-wavelength shells and their outflows collide almost instantaneously, efficiently thermalizing their kinetic energy. Additionally, the perpendicular outflows carry away considerable amounts of kinetic energy, with thermal energy being somewhat more concentrated towards the jet axis, leading to a further apparent increase in internal energy deposition efficiency as defined in this work. At frequencies above the minimum, $\omega_2 > 0.25$, the roles of the two components reverse. Now the low-amplitude long-wavelength (second) component is modulating the high-amplitude short-wavelength (first) component. This leads to more efficient thermalization following the same mechanism. The high- and low-frequency behavior will also reverse for $\Gamma_{\rm max, 2} < 5$, as the second component becomes the low-amplitude one, and the first component with $\Gamma_{\rm max, 1} = 5$ has higher amplitude.

It is interesting to note that the position of the minimum is not at $\omega_2 = 2$, where the two patterns would have the same frequency. This is likely related to the fact that the two components are injected in-phase. As the two components approach each other in frequency, the model simplifies to a single-component model with a significantly larger maximal Lorentz factor. This resonance greatly improves thermalization efficiency despite the fact that the collisions between shells, aiding internal energy deposition in other regions of the parameter space, no longer happen. As a result, the minimum in thermalization efficiency is moved from $\omega_2 = 2$, $\Gamma_{\rm max, 2} = 5$, where it would be if shell collisions were the only relevant factor, to lower $\Gamma_{\rm max, 2}$, where the shell collisions are still relevant, but the two components are far enough from resonance. 

\begin{figure}
 \centering
 \includegraphics[resolution=600]{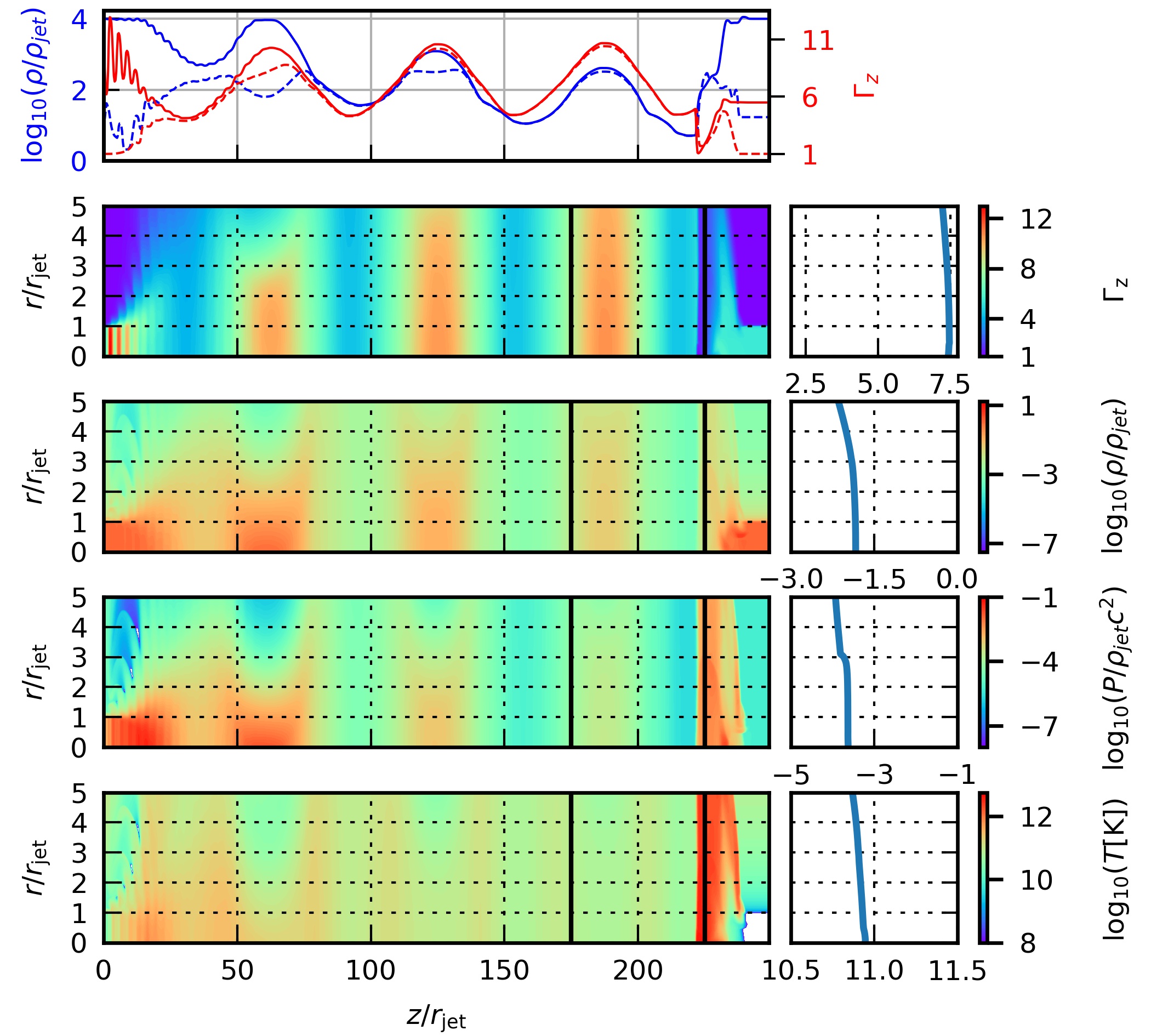} \includegraphics[resolution=600]{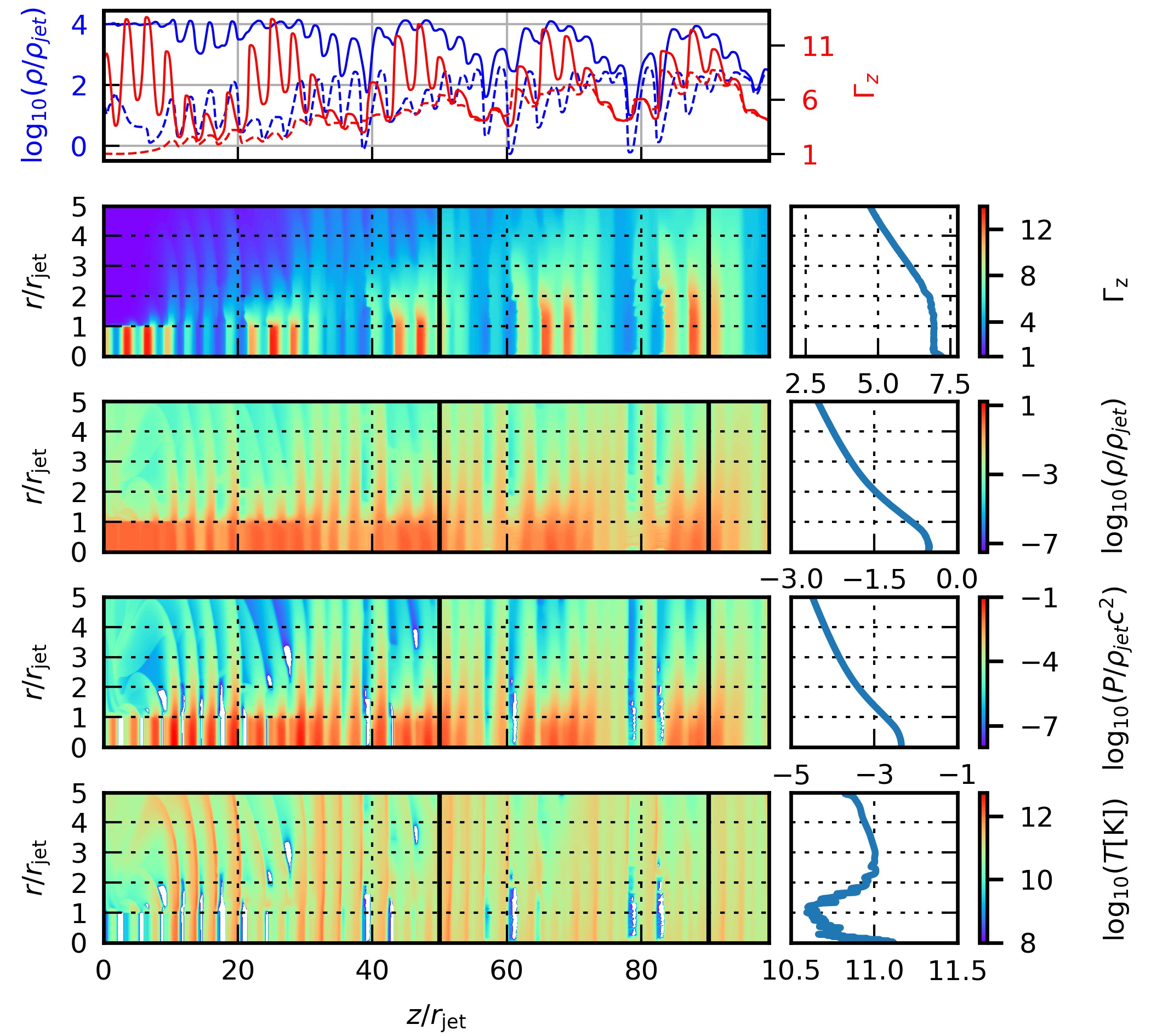}
 \caption{Snapshots at different injection frequencies surrounding the minimal efficiency case for two-component injection (top: $\omega_2 = 0.1$, $t=235.6$~sim.u.; bottom: $\omega_2 = 0.3$, $t=105.8$~sim.u.). In each case $\Gamma_{\rm max, 1} = 10$ and $\Gamma_{\rm max, 2}=8.09$ so that the latter frame of reference lags behind $\Gamma_{\rm max, 1}$ at a (relative) Lorentz factor of $\Gamma_{\rm rel} = 3$. The meaning of symbols and notation are the same as in Fig.~\ref{fig:qualitative}.}
 \label{fig:Two-EdepExtNu}
\end{figure}

\begin{figure}
 \centering
 \includegraphics[resolution=600]{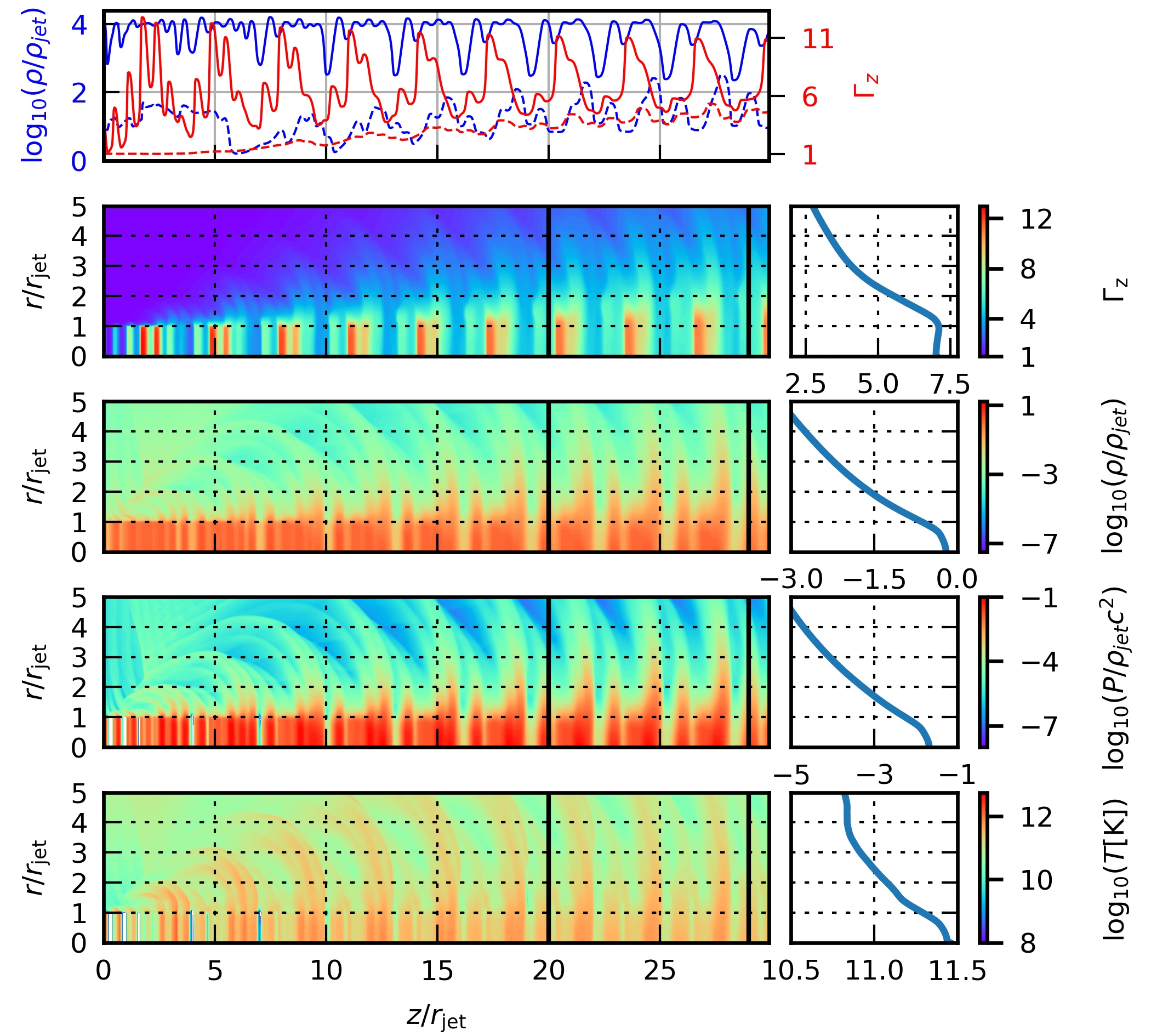}
 \caption{Continuation of Fig.~\ref{fig:Two-EdepExtNu}, a snapshot for $\omega_2 = 10$, $t=50.5$~sim.u., $\Gamma_{\rm max, 1} = 10$, $\Gamma_{\rm max, 2}=8.09$ ($\Gamma_{\rm rel} = 3$). The meaning of symbols and notation are the same as in Fig.~\ref{fig:qualitative}.}
 \label{fig:Two-EdepExtNu2}
\end{figure}

\begin{figure}
 \centering
 \includegraphics[resolution=600]{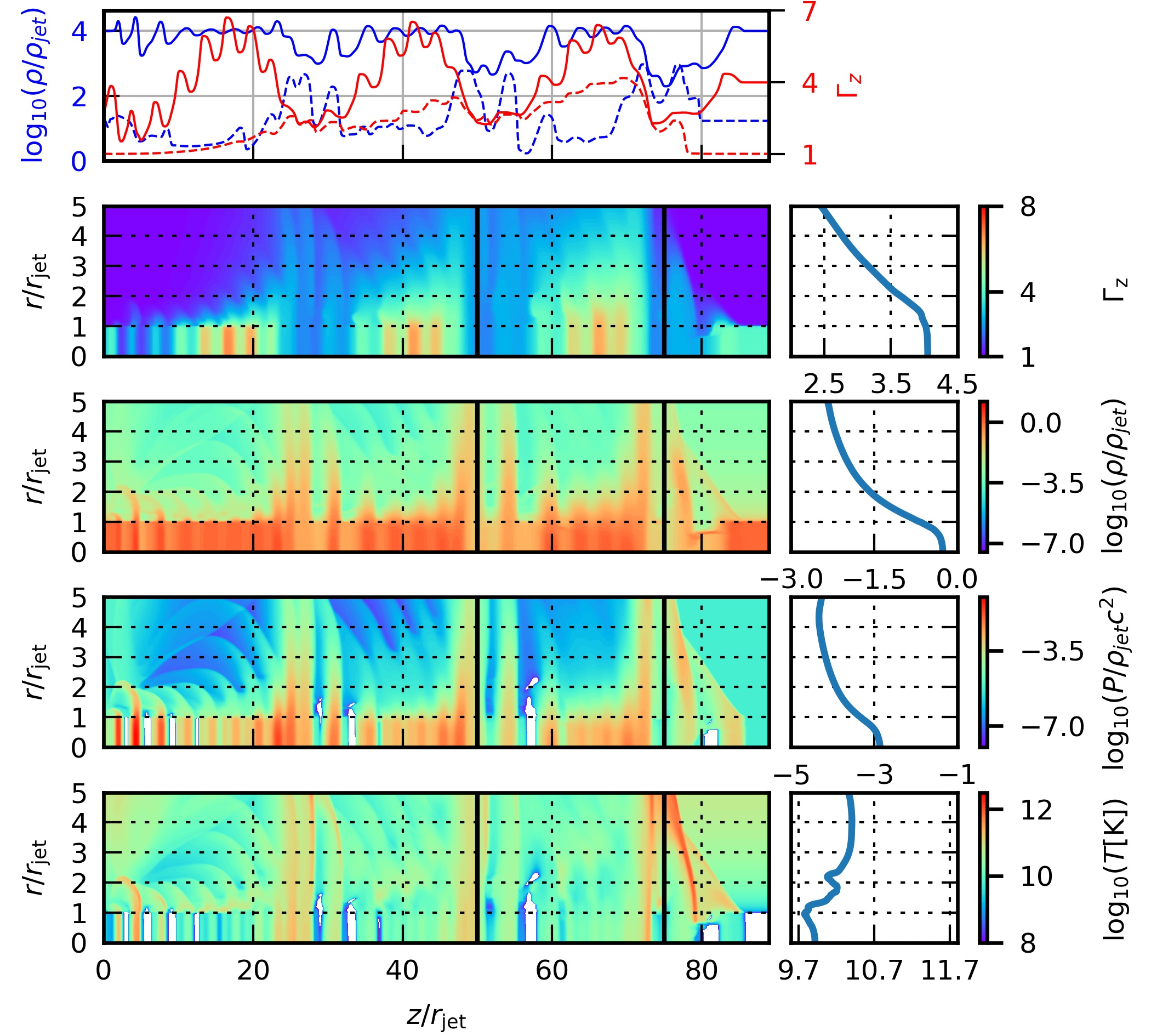}\\
 \includegraphics[resolution=600]{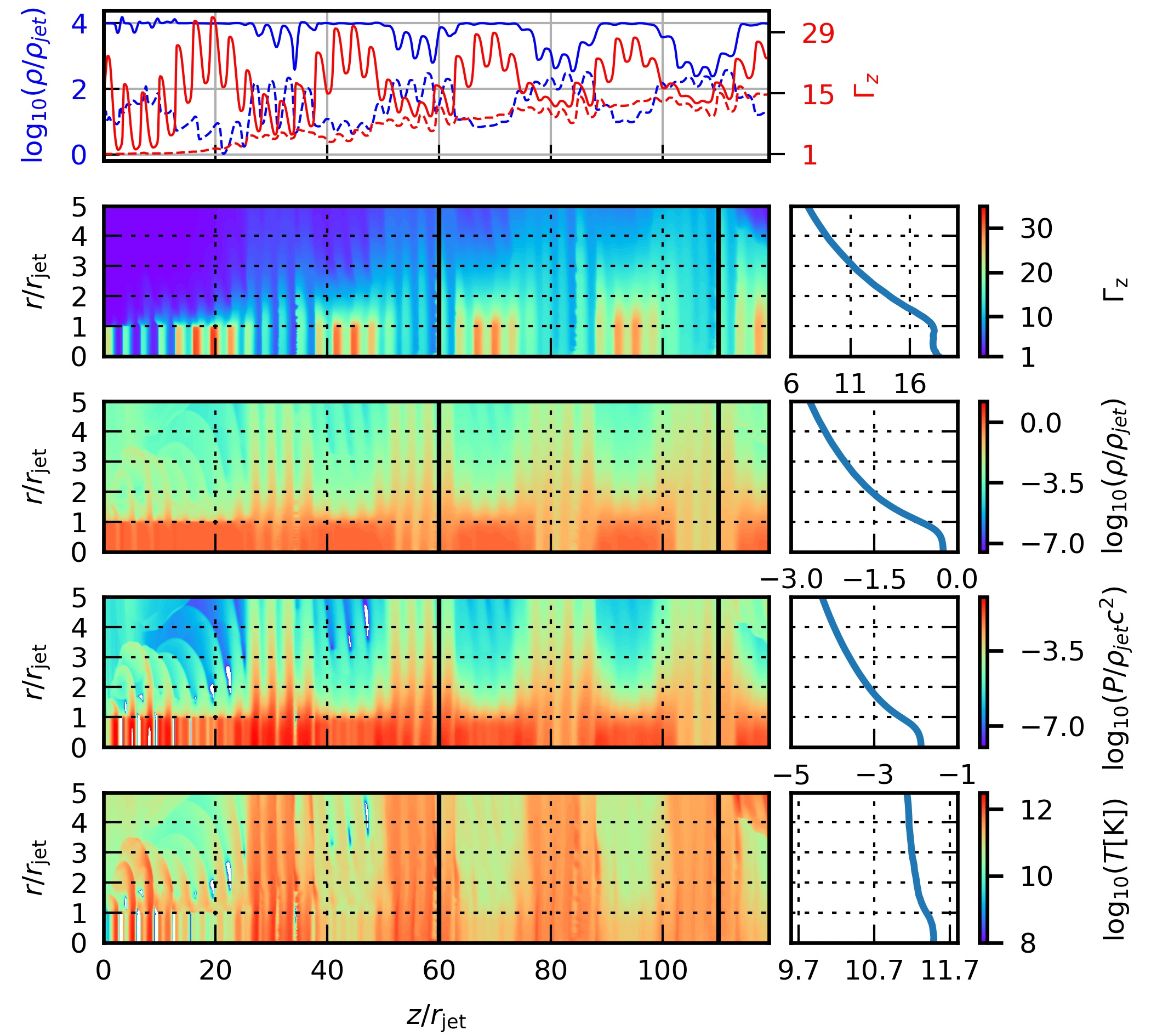}
 \caption{A snapshot of two extremal-$\Gamma_{\rm max, 2}$ cases (top: $\Gamma_{\rm max, 2} = 2$, $t=88.0$~sim.u.; bottom: $\Gamma_{\rm max, 2} = 25$, $t=126.9$~sim.u.) for two-component simulations with $\omega_2 = 0.25$, $\omega_1 = 2$, $\Gamma_{\rm max, 1} = 5$. The meaning of symbols and notation are the same as in Fig.~\ref{fig:qualitative}.}
 \label{fig:Two-EdepExtGam}
\end{figure}

\subsubsection{Maximum temperature as a function of the relative speed of the pattern components}

The maximum temperature contours seem to be much better centered around the expected case of the two pattern components becoming degenerate ($\omega_2 = 2$, $\Gamma_{\rm max, 2} = 5$), at which position it has a minimum (Fig.~\ref{fig:2shell:Eth_contour}). It therefore appears that the temperature is relatively insensitive to increased Lorentz factors in resonance between the two pattern components, and instead follows the strength of the shocks between colliding shells. This suggests that the shell collisions are more efficient at increasing the maximal temperature reached than the increase the Lorentz factor modulation amplitude.

\subsubsection{Mach numbers of the shocks}

The behavior of Mach numbers of the shocks in two-component pattern injection case is qualitatively different from that of a single sinusoid injection. Their values are governed by interference of the two components -- both within a single simulation run (where the Mach number of a shock varies as the shocks collide and merge) and across the tested parameter space. In the latter case, Mach numbers are most sensitive to the difference in frequency between the two components, with extremely high Mach numbers (up to $10^8$) being achieved when the two components have almost the same frequency. At this point, the two components merge and form an effectively single component with a very high amplitude, which evacuates the inter-shell regions very efficiently.

Both the reverse and forward shock Mach numbers decay with time for two-component modulation models. While maximum values of the forward shock Mach numbers are much higher than in the single-component case, they eventually reach the same range of $10^4$-$10^5$. Reverse shock Mach numbers remain within this range for the entire simulation.

\subsection{Interaction of multiple components\\ -- custom PSD injection}\label{sect:somanyshells}

\begin{figure*}
 \centering
 \includegraphics{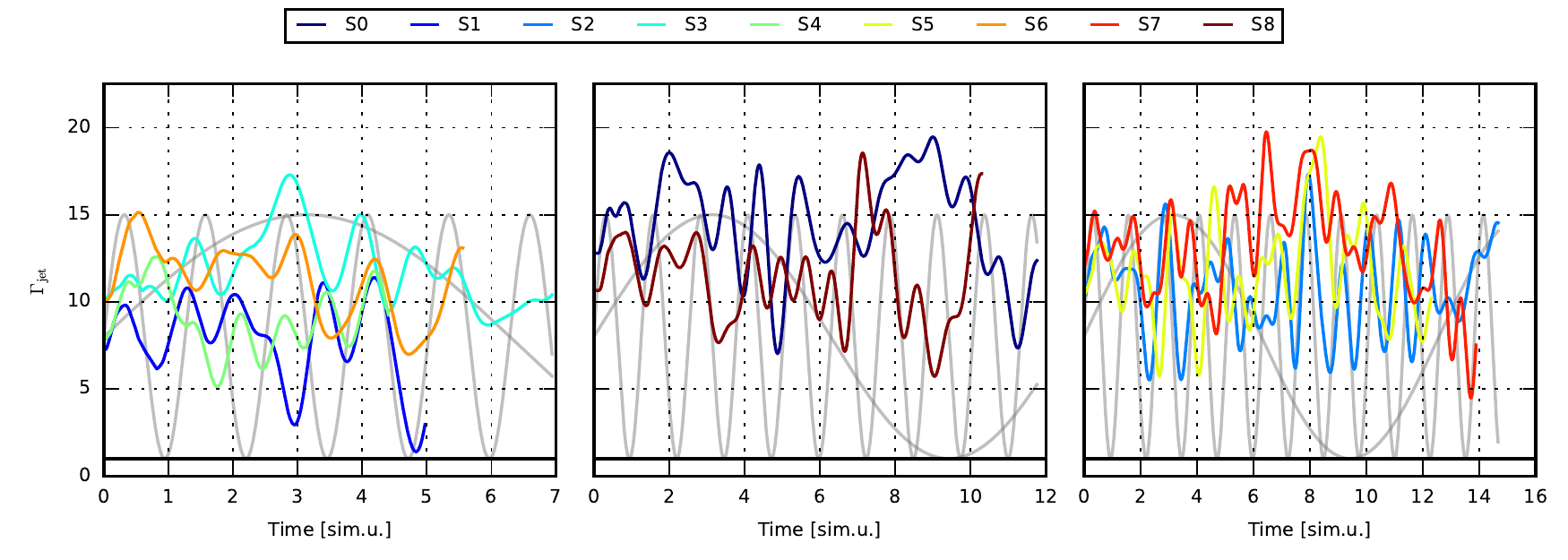}
 \caption{Combined shell patterns for the nine simulations with a flicker-noise-spectrum multiple component injection (see Sect.~\ref{sect:shellInjection}). Each pattern is shown over one wavelength of its longest-wavelength sinusoidal component. Note that, as each pattern's components are incommensurate, the continued combined pattern will differ from the fragment shown. The patterns are divided into three bins by the wavelength of their lowest-frequency component for clarity. The grey lines on the plots show the shell patterns of two single component simulations with $\Gamma_{\rm max} = 15$, $\rho_{\rm jet}/\rho_{\rm amb} = 10^4$, $\omega = 0.5$ and $\omega = 5$, respectively, for comparison with the multiple component results. The black horizontal solid line marks the position of null velocity.}
 \label{fig:patterns_nshell}
\end{figure*}

Finally, we follow with simulations including multiple components of the Lorentz factor modulation. As described in Sect.~\ref{sect:shellInjection}, in these cases $20$ components are chosen with random $\omega\in [0.1,10]$, amplitudes $\Delta\Gamma\in [0.5,5.0]$ following $\Delta\Gamma \propto \omega^{-1/2}$, and random phases. The pattern is then shifted so that its minimum Lorentz factor corresponds to null velocity. The resulting patterns for the $9$ simulations performed are shown in Fig.~\ref{fig:patterns_nshell}. Here, we will compare them to the single-shell propagation simulations with $\Gamma_{\rm max} = 15$ and $\omega = 0.5, 5.0$, which roughly follow the dominant components of most patterns.

\begin{figure*}
 \centering
 \includegraphics{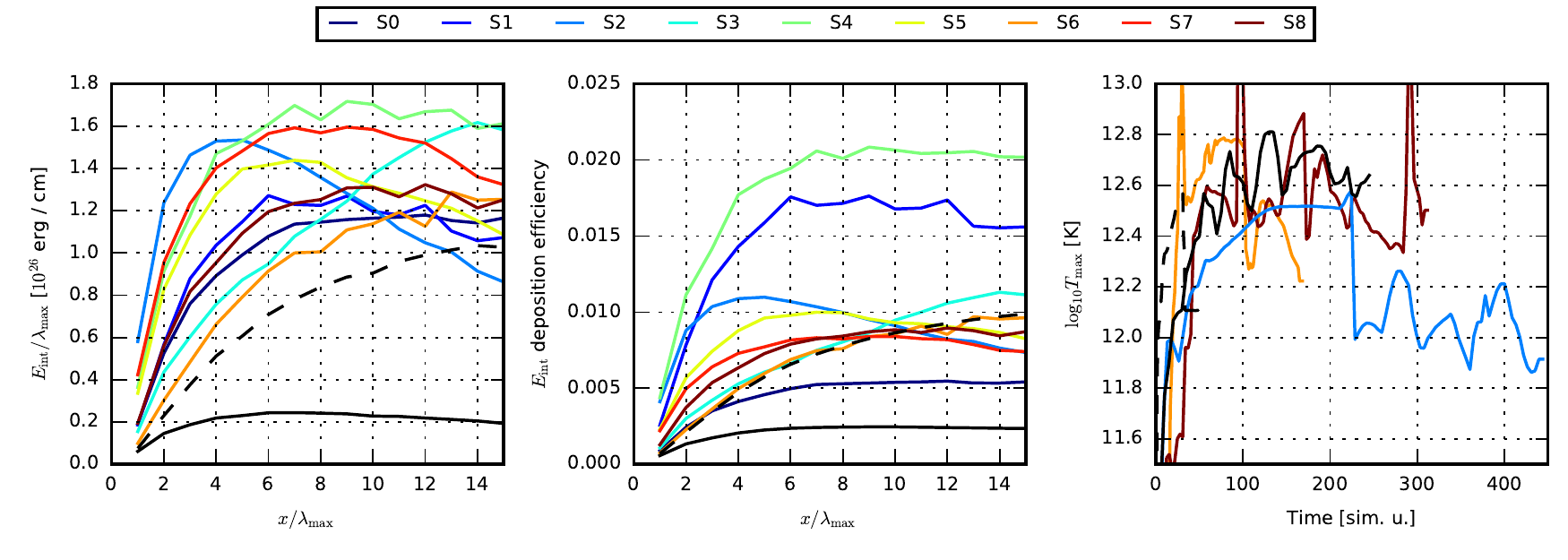}
 \caption{Results of the multiple-component shell pattern injection simulations. First column: the internal energy deposited per wavelength of the lowest-frequency component in the pattern as a function of the integrated area position (see Sect.~\ref{sect:diagnostics} for details). Second column: internal energy deposition efficiency as a function of position downstream the jet. In the first two columns the solid and dashed black lines show the corresponding dependencies for single-component simulations with $\Gamma_{\rm max} = 15$, $\rho_{\rm jet}/\rho_{\rm amb} = 10^4$, $\omega = 0.5$ and $\omega = 5$, respectively, with results (deposited energy and efficiency) decreased by a factor of ten. Third column: maximum temperature in the simulation as a function of time (tracks of only three representative multiple-shell simulations are shown for clarity). The solid and dashed black lines in the right column show $T_{\rm max}$ for single shell simulations with $\Gamma_{\rm max} = 15$, $\rho_{\rm jet}/\rho_{\rm amb} = 10^4$, $\omega = 0.5$ and $\omega = 5$, respectively. The color designations of simulation runs are the same as in Fig.~\ref{fig:patterns_nshell}.}
 \label{fig:res_nshell}
\end{figure*}

All multiple-component models have lower deposited energies and deposition efficiencies than the two single-component cases (Fig.~\ref{fig:res_nshell}), despite the maximal Lorentz factors of the latter being lower. One might expect that this is because the Lorentz factor variability, not its absolute value, sets the amount of kinetic energy available for thermalization. That being said, we find no correlation between either the amount of the deposited internal energy or the deposition efficiency and the standard deviation of the injected shell pattern. We expect that these differences result from the interplay between the shells and outflows at different speeds in a complex pattern. At present, we lack a good diagnostic to simplify this relation.

Multiple-shell simulations exhibit a range of behaviors both in the total deposited energy and the deposition efficiency. Some reach a plateau after several wavelengths (e.g., S0) from the injection point. Others efficiently deposit most of their energy early and deposit less internal energy in each following wavelength (e.g., S2). Others yet slowly increase the amount of the internal energy deposited, with the trend continuing even after 15 wavelengths (S3).
We attribute this behavior to jet -- ambient medium mixing in these three cases. In the case of simulation S0, the shell pattern (see Fig.~\ref{fig:patterns_nshell}) is composed of sections with long- and short-wavelength dominant behavior. The short-wavelength shells collide and disperse early after injection, ejecting some perpendicular outflows, but without mixing the jet with the ambient medium efficiently. Then, the long-wavelength variations launch their outflows into a relatively pristine ambient medium, which allows them to draw strong shocks into the medium and support efficient thermalization at later times. Simulation S2, on the other hand, is dominated by short-wavelength variability. The shells form early and eject perpendicular outflows, shocking the ambient medium and supporting efficient thermalization. The outflows mix the jet with the ambient gas, preventing strong shocks from forming later on. Moreover, small-size shells disperse into each other, preventing additional outflows from being launched. This causes the internal energy deposition rate to decrease downstream the jet. We note that the short-wavelength single-shell simulation ($\omega=5$, dotted curve in Fig.~\ref{fig:res_nshell}) also has a peaked deposited energy distance dependence. Finally, in the case of S3, the dominant behavior is long-wavelength (Fig.~\ref{fig:patterns_nshell}). The shells and outflows evolve more slowly, allowing internal energy deposition to increase over multiple wavelengths.
The long-wavelength single-shell simulation ($\omega=0.5$) also follows this route (dashed curve in Fig.~\ref{fig:res_nshell}).

While the deposition efficiency shows a range of behaviors dependent on the specific random set of component parameters, most of the cases cluster around the deposition efficiency of $\sim 8\%$. This value is in agreement with similar studies performed for GRBs, e.g., \cite{2009Maxham} and \cite{2015Gao}, who achieve $\eta_{\rm dep} \sim 10\%$ for collision parameters similar to ours. It is, however, significantly lower than the result for a single collision between magnetized shells in full MHD, which is reported by \cite{2015Deng} to reach $\sim 36\%$ (see Sect.~\ref{sect:magnetic_fields} for further discussion). Simulations S0, S1 and S4 exhibit deposition efficiencies significantly different from those of other simulation runs. Pattern S0 contains large regions of low-variability in $\Gamma_{\rm jet}$ (see Fig.~\ref{fig:patterns_nshell}), decreasing the number of shocks produced. Moreover, its Lorentz factor remains high for much of the pattern\footnote{Note that Fig.~\ref{fig:patterns_nshell} only shows the first wavelength of the lowest-frequency component of each pattern. As the frequencies of its components are incommensurate, the shape of the pattern will continue to evolve at later times.}, preventing shocks from forming within the jet and powerful outflows from being launched. As a result, its deposition efficiency is low. In contrast, simulations S1 and S4 are highly variable with the Lorentz factor reaching very low values frequently throughout the pattern. This results in their high efficiency.

The maximum temperatures reached in multiple-component simulations are similar to those of both single- and two-component simulations and fluctuate around $\sim 10^{12}$~K. $T_{\rm max}$ does, however, exhibit significant drops for longer simulations, e.g., S7. This is likely an effect of mixing of the material, which causes the ambient medium shocks, where $T_{\rm max}$ is usually reached, to be weaker. As a shell with strong shocks leaves the simulation box, the maximum temperature is set by the remaining shells. As their perpendicular outflows not being able to shock the pre-mixed ambient medium as strongly, this temperature is lower.

Due to the complicated nature of shock interactions in the case of multiple-component Lorentz factor modulation, we were unable to reliably establish their Mach numbers. Manual inspection of a few selected shocks hints that the Mach numbers here should be similar to those in the two-shell models.

\subsubsection{Cyg~X-1}

Throughout this work we have used a model of Cyg~X-1 from \cite{2014Zdziarski} to anchor our simulation in physical units. While we stress that a modelling of this system is not the goal of this work, and that the physical units reported are only intended to give an order-of-magnitude intuition to the actual values we might expect, it is still interesting to attempt to use our work with respect to Cyg~X-1. Typical X-ray variability spectra for this high mass X-ray binary can be found, e.g., in fig.~2 of~\cite{2005Axelsson}. In the hard state of the system, where a steady jet emission is observed, the PDS of the fluctuations in the $\log(f\times \textrm{PDS})$~vs~$\log f$ space (where $f$ denotes frequency) is well modelled by a sum of two Lorentzians forming a plateau between the peak frequencies. Most of the variability power is therefore contained as flicker-noise variability between these two frequencies. The positions of the two peaks vary considerably (see fig.~9 of \citealt{2005Axelsson}), but $\nu_1 \sim 0.5$~Hz and $\nu_2\sim 5$~Hz provide good typical values. These translate to $\nu_1 \sim 0.001$ and $\nu_2\sim 0.010$ in our simulation units. We readily note that these frequencies are well below any modulation frequencies we have tested in this manuscript. In order to resolve the wavelengths of modulations at these frequencies, the aspect ratio of the simulation box would have to be very large. Since the time step would be given by the size of each cell perpendicular to the jet axis (which must remain small to resolve the jet itself), performing such simulations would be computationally very expensive. While we need to keep this in mind, let us assume that the $\sim 9\%$ thermalization efficiency measured for our multiple-component simulations holds for lower frequencies than tested. Let us estimate the amount of thermal energy available to Cyg~X-1 hard-state jet under this assumption. The average bulk Lorentz factor of this jet is rather low, $\Gamma_j \sim 1.5$ \citep{2001Stirling, 2004Gleissner, 2009Malzac}, so we assume the modulation pattern to generate Lorentz factors between $1.0$ and $2.0$. Kinetic energy associated with the relative motions of the jet fluid in such a case should be of the order of $0.5M_jc^2$, where $M_j$ is the total mass of the jet. The amount of internal energy generated by internal shells will then be $\sim 0.05M_jc^2$. Assuming that the entire accreted mass is transferred into the jet, $\dot{M}_j \sim \dot{M}_{\rm acc}$, the ``thermalization power'' for the jet will be $\sim 0.05\dot{M}_{\rm acc}c^2$. Given that the radiative efficiency of the Model~1 of \cite{2014Zdziarski} is $\sim 0.003$, we come to a conclusion that, under our assumptions, $\sim6\%$ of the kinetic energy dissipated in shocks we describe would have to be used for electron re-acceleration in order to explain the observed synchrotron emission.

\subsection{Caveats}

\subsubsection{Lateral expansion}

While we simulate the jet as a cylindrical flow, true jets are of course much better approximated by conical outflows. This fact adds an additional sink to the energy balance, causing part of the internal energy of the jet to be consumed by the adiabatic lateral expansion of the jet. This lowers the pressure effectively generated between shocks bounding the internal shells and causes the perpendicular outflows to be less powerful, resulting in less efficient thermalization of the jet bulk kinetic energy.

\subsubsection{Magnetic fields}\label{sect:magnetic_fields}

A major caveat of our treatment of a jet is the lack of magnetic fields. In our target case of an X-ray binary jet section far from the ejection region, the magnetization of the jet is expected to be small \citep[see][and references therein]{2015Zdziarski}. Magnetic fields are therefore not dynamically important and should not affect the hydrodynamic effects described here significantly. As indicated by non-relativistic MHD simulations of pulsed protostellar jets performed by \cite{2000Stone}, a combination of hoop stress and magnetic pressure will likely result in less powerful perpendicular outflows as the magnetic field is increased.

Magnetic fields can, however, be extremely important close to the ejection site of the jet, where the magnetization is likely high and magnetic fields can be dynamically important \citep[e.g.,][]{2012McKinney}. In such an environment, magnetic fields could prevent the ejection of perpendicular outflows, which drive thermalization in the hydrodynamical case described here. In addition, if the field is toroidal (as indicated by observations, e.g., \citealt{2014Russell}), it can also prevent internal shells from forming as it would transfer the kinetic energy of Lorentz factor modulation into magnetic pressure. On the other hand, strong magnetic fields can lead to reconnection, which can be extremely efficient in thermalizing relative motion of magnetized blobs in the jet, as shown by \cite{2015Deng}, who achieve $35\%$ thermalization efficiency for the energy stored in the magnetic field. We therefore caution that our treatment is not appropriate for use in high-magnetization regions of jets.

\section{Conclusions}\label{sect:discussion}

We performed a set of hydrodynamic simulations of a relativistic jet with continuous modulation of the bulk Lorentz factor. We have investigated modulation in form of a single sinusoidal component, as well as a sum of two and $20$ components. Our diagnostics included the internal energy deposited within three jet radii from the jet axis per unit length along the jet, ``efficiency'' of this deposition (defined as the ratio of internal to total flow energy), and the maximum temperature reached in each simulation. We find that, at each peak of the injected Lorentz factor variation pattern, the modulation produces a forward-and-reverse shock structure, not dissimilar to that found in supernova remnants. These shocks enclose a high-density, high-temperature region (an internal shell), which launches powerful outflows into the ambient medium perpendicularly to the jet axis. We find the mixing of jet and ambient matter facilitated by these outflows and their shocking of the ambient medium to be the key factors driving thermalization in our models. We also note that the interaction between the outflows themselves affect the internal energy deposition, as each outflow pre-shocks the ambient medium encountered by the consecutive one.

For the single-component modulation models, we find the following dependencies between model parameters and our diagnostics:
\begin{enumerate}
 \item The density contrasts only affect the results of our simulations when they are low, $\rho_{\rm jet}/\rho_{\rm amb} \lesssim 50$. At these values, the deposition efficiency increases with decreasing density contrast due to the ambient medium quickly evacuating the internal shells of pressurized gas.
 \item At low density contrasts, the internal energy deposition is insensitive to the Lorentz factor modulation frequency.
 \item At high density contrasts, increasing modulation frequency leads to increased internal energy deposition. We attribute this effect to jet disruption caused by the ``shielding effect'' perpendicular outflows have on each other.
 \item At high density contrasts, the internal energy deposition efficiency is maximized at a certain Lorentz factor variation amplitude at each modulation frequency. The amplitude at which the maximum occurs is higher for increasing modulation frequencies. Emergence of the maximum can be explained by the ability of the model to form well-defined shells, limited on one side by ultra-relativistic character of the peaks and on the other by the relative motion of the forward and reverse shocks.
 \item The maximum temperatures reached in our simulations remain remarkably stable and are always within an order of magnitude of $10^{11.9}$~K. We find that they are always reached in the shocks driven by the perpendicular outflows into the ambient medium (which is initialized at low density and high pressure, and thus already hot).
 \item The maximum temperature remains sensitive to density contrasts even at very low relative densities of the ambient medium.
\end{enumerate}

Addition of a second sinusoidal component to the Lorentz factor modulation expands our understanding of variable jets further:
\begin{enumerate}
 \item Both the deposited internal energy and deposition efficiency have a minimum close to where the two pattern components are identical, suggesting that the collisions between shells are an important mechanism of thermalization in these models. The influence of collision-driven thermalization is maximized when the frequencies of the two pattern components are significantly different.
 \item The internal energy deposition and its efficiency are more efficiently amplified through Lorentz factor increase at resonance between the two components than through the shell collisions away from resonance. However, the shocks produced by shell collisions reach higher maximal temperatures than those from interaction of single-component outflows with the ambient medium, even at higher Lorentz factors caused by resonance between the pattern components.
\end{enumerate}

We also investigate a case of flicker-noise Lorentz factor variations, which \cite{2014Malzac} found to be able to explain the flat radio spectra of the observed relativistic jets. We approximate these variations by a sum of $20$ sinusoidal components of the Lorentz factor modulation. For these models we find that:
\begin{enumerate}
 \item The deposition efficiencies are about an order of magnitude lower than in the case of single-component modulation with comparable amplitudes and dominant frequencies. While this is likely a result of the difference in the range of Lorentz factors reached in these two types of models, we find no correlation between the internal energy deposition efficiency and the Lorentz factor standard deviation between the individual multiple-component patterns.
 \item The distance from the modulation injection point over which the jet bulk kinetic energy is converted into internal energy varies between specific realizations of the PSD, depending on the dominant frequencies in a specific pattern. Larger samples of the PSD may shed more light on this behavior in future studies. Overall, the deposition efficiency for the multiple-component cases clusters around $8\%$.
\end{enumerate}

\section*{Acknowledgements}

Computations were performed on computer systems provided by the
Princeton Institute for Computational Science and Engineering.

We would like to thank the anonymous reviewer for helpful comments that improved this manuscript.




\bibliographystyle{mnras}
\bibliography{references} 






\bsp	
\label{lastpage}
\end{document}